\documentclass[journal=jpclcd,manuscript=article]{achemso}
\usepackage[version=3]{mhchem}
\usepackage{commath}
\usepackage{graphicx}
\usepackage{subcaption}
\usepackage{multirow}
\usepackage{booktabs}
\usepackage{silence}
\usepackage{array}
\usepackage{braket}
\usepackage[T1]{fontenc}

\author{Jiachen Li}
\affiliation{Department of Chemistry, Duke University, Durham, NC 27708, USA}
\author{Zehua Chen}
\affiliation{Department of Chemistry, Duke University, Durham, NC 27708, USA}
\author{Weitao Yang}
\affiliation{Department of Chemistry, Duke University, Durham, NC 27708, USA}
\email{weitao.yang@duke.edu}

\title{Renormalized Singles Green's Function in the T-Matrix Approximation for
Accurate Quasiparticle Energy Calculation}

\begin{document}

\begin{tocentry}

\includegraphics[width=1\textwidth]{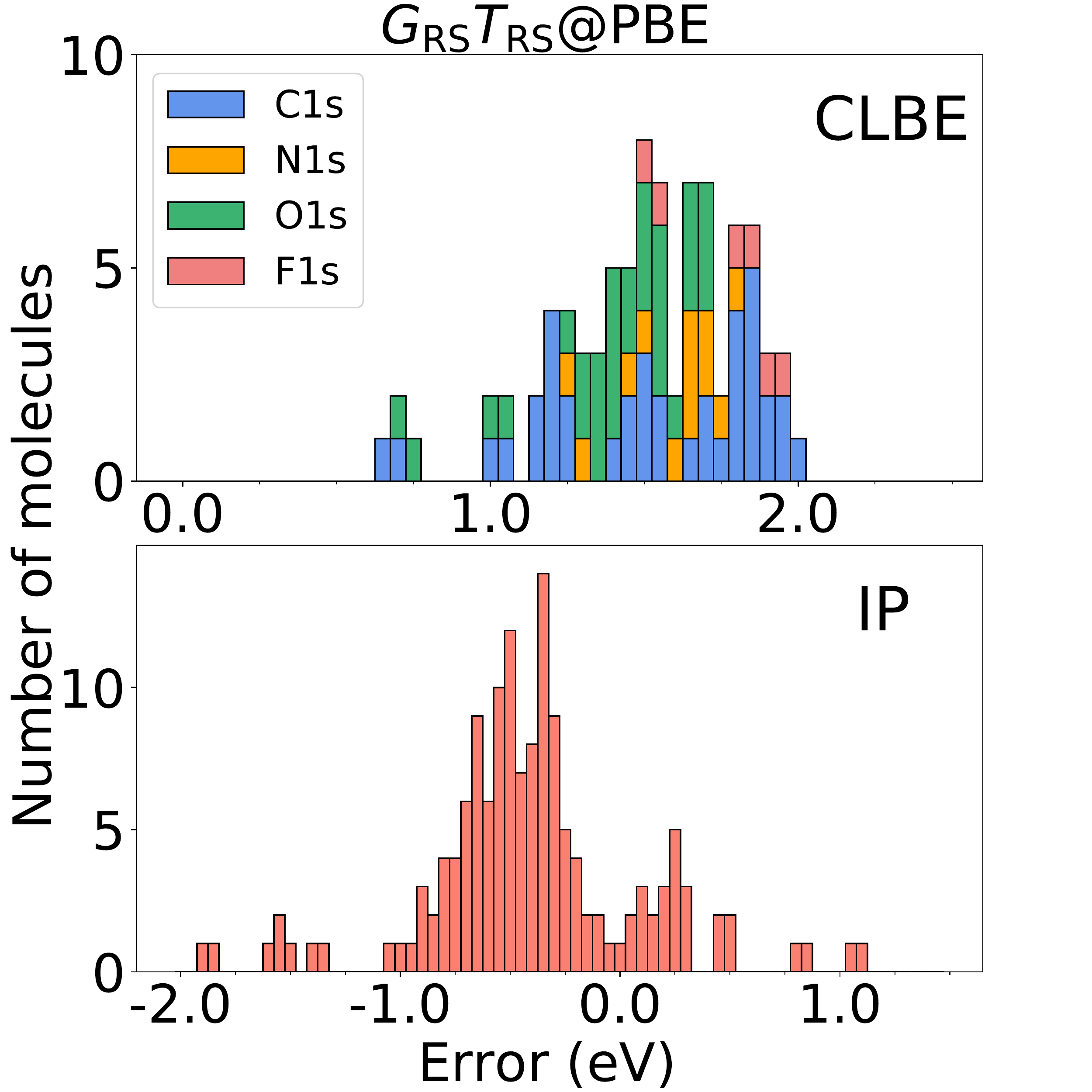}

\end{tocentry}

\begin{abstract}
We combine the renormalized singles (RS) Green's function with the T-Matrix
approximation for the single-particle Green's function to compute quasiparticle
energies for valence and core states of molecular systems. The
$G_{\text{RS}}T_0$ method uses the RS Green's function that incorporates singles
contributions as the initial Green's function. The $G_{\text{RS}}T_{\text{RS}}$
method further calculates the generalized effective interaction with the RS
Green's function by using RS eigenvalues in the T-Matrix calculation through
the particle-particle random phase approximation.  The
$G_{\text{RS}}T_{\text{RS}}$ method provides significant improvements over the
one-shot T-Matrix method $G_0T_0$ as demonstrated in calculations for GW100 and
CORE65 test sets. It also systematically eliminates the dependence of
$G_{0}T_{0}$ on the choice of density functional approximations (DFAs). For
valence states, the $G_{\text{RS}}T_{\text{RS}}$ method provides an excellent
accuracy, which is better than $G_0T_0$ with Hartree-Fock (HF) or other DFAs.
For core states, the $G_{\text{RS}}T_{\text{RS}}$ method correctly identifies
desired peaks in the spectral function and significantly outperforms $G_0T_0$
on core level binding energies (CLBEs) and relative CLBEs, with any commonly
used DFAs.
\end{abstract}

Quasiparticle (QP) energies constitute one of the most important electronic
properties of molecules and materials. Although they can be measured in
photoemission and inverse photoemission spectroscopies, the computational study
of QP energies plays an important role for understanding electronic structures
from basic principles and for molecular and material designs.
The $GW$ method\cite{martin2016interacting, reining2018gw, golze2019gw},
which is developed from Hedin's equations\cite{hedin1965new}, is a state-of-art
formalism to study QP energies, and charged electronic excitations in molecules\cite{ke2011all, wilhelm2018toward, maggio2017gw, ren2021all, caruso2012unified, wilhelm2021low,caruso2014first, hellgren2015static}
and solids\cite{rinke2005combining,rinke2009defect, marom2017accurate, jiang2010first, rinke2006band, bakhsh2021beryllium, trevisanutto2008ab, aguilera2011first, pulci1999state}.
The success of the $GW$ method stems from a clear physical interpretation for
QP energies, a favorable computational scaling with respect to the size of
systems and a proper description of the screened
interaction\cite{martin2016interacting}. The $GW$ method is viewed
as the "gold standard" for band gap calculations\cite{blase2018bethe} for
periodic systems, and the $GW$ method has been widely used to investigate both
valence\cite{van2015gw, caruso2016benchmark, knight2016accurate} and
core\cite{golze2019accurate, zhou2015dynamical, golze2018core} state properties
for molecular systems. Despite enormous successes achieved by $GW$ in practical
calculations, the most commonly-used $GW$ variant, the one-shot method
$G_0W_0$, still suffers from the strong starting-point dependence. In practice,
the $G_0W_0$ method is usually combined with a preceding Kohn-Sham (KS) density
functional theory (DFT) calculation\cite{hohenberg1964inhomogeneous, kohn1965self, parr1989density}
calculation. The difference originated from using different density functional
approximations (DFAs) can exceed 1 eV for ionization potentials (IPs) and
electron affinities (EAs) of molecules\cite{marom2012benchmark, ke2011all} and
can be even larger than 2 eV for binding energies of solids\cite{fuchs2007quasiparticle}.
Partial or full self-consistent $GW$ approaches, such as ev$GW$ and sc$GW$
can greatly improve the accuracy and eliminate the starting-point dependence
for both valence and core state calculations\cite{kaplan2016quasi,caruso2016benchmark, caruso2013self},
but inevitably bring additional computational cost. \par

Another approach in $GW$ is using the (RS) Green's function\cite{jin2019renormalized}
as the new starting point, which is denoted as $G_{\text{RS}}W_{0}$ and
$G_{\text{RS}}W_{\text{RS}}$\cite{li2021core}. The motivation of the RS Green's
function is to use the form of the Hartree-Fock\cite{szabo2012modern, slater1930note}
(HF) self-energy to include singles contributions because it captures all the
contributions of the single excitations completely, unlike in the commonly used
$G_{0}W_{0}$, which is done perturbatively. The RS as the reference
noninteracting single-particle Green's function shares similar thinking as the
renormalized singles correction for correlation energies\cite{ren2011beyond, ren2013renormalized}.
The HF Green's function itself is not a good starting point for $G_0W_0$\cite{bruneval2013benchmarking, jin2019renormalized}.
However, the renormalization process of forming the HF self-energy with DFA
orbitals significantly improves the accuracy and eliminates the dependence on
orbital energies of the DFA\cite{jin2019renormalized}. Furthermore, the
incorporation of singles contributions, using RS eigenvalues with KS orbitals
has been shown to lead to corrections to the underestimation of excitation
energies from the particle-hole random phase approximation (phRPA) based on
commonly used KS DFA\cite{peng2014restricted}. The underestimated excitation energies from the phRPA
at DFA levels erroneously transfer spectral weights from desired QP peaks to
satellites\cite{golze2020accurate}.
The improved phRPA excitation energies lead to a unique solution of the QP
equation in the core region\cite{li2021core}. The renormalization step is a
much less computational-demanding step than solving the phRPA equation and
formulating the self-energy. Developed in the first paper on the RS Green's
function, the $G_{\text{RS}}W_{0}$ method uses the RS Green's function as the
new starting point but calculates the screened interaction with the KS Green's
function and it leads to considerable improvements over $G_0W_0$ for IPs and EAs\cite{jin2019renormalized}.
Then the $G_{\text{RS}}W_{\text{RS}}$ method that also uses the RS Green's
function in the screened interaction has been shown to provide better results
than $G_0W_0$ for core level binding energies (CLBEs) and it reproduces correct
behaviors for spectral functions\cite{li2021core}. \par

In this paper, we introduce the RS Green's function to the counterpart of $GW$
in the particle-particle (pp) channel, which is the T-Matrix method. The
T-Matrix approximation or the Bethe-Goldstone approximation was originally
introduced to treat the interaction of complex nuclei\cite{bethe1957effect,baym1961conservation,baym1962self,danielewicz1984quantum,danielewicz1984quantuma}.
The T-Matrix method was used to describe electronic structures of the
Hubbard model\cite{bickers1989conserving,bickers1991conserving,von2010kadanoff,gukelberger2015dangers, romaniello2012beyond, romaniello2012beyond}.
It has also been applied to study periodic systems in material science,
including satellites\cite{springer1998first, guzzo2011valence}, double excitations\cite{zhukov2005gw+, noguchi2010cluster}
and spin-flip excitations of metals\cite{muller2019electron, mlynczak2019kink}.
It has been shown that the T-Matrix method works well for the low-density
limit, where the screening effect has less impact and the $GW$ approximation
fails\cite{danielewicz1984quantum,liebsch1981ni}. \par

Recently, the T-Matrix method was introduced to predict accurate QP spectra of
molecules\cite{zhang2017accurate}, motivated by desirable improvements on
predicting excitation energies of the particle-particle random phase
approximation (ppRPA)\cite{van2013exchange,van2014exchange,yang2013benchmark}
over the phRPA\cite{bohm1951collective,bohm1953collective}. In the T-Matrix
approximation, the self-energy is formulated with a four-point generalized
effective interaction $T(1,3;2,4)$ instead of the two-point screened
interaction $W(1,2)$. The advantage of using the four-point generalized
effective interaction is that, the self-energy includes electron exchange
interactions and it is exact up to second order in the bare interaction, while
in the $GW$ method the self-energy does not have these properties. Similar to
the $GW$ approximation that is based on the phRPA, the T-Matrix approximation
has been formulated with the pairing excitation eigenvalues and eigenvectors of
the ppRPA matrix\cite{zhang2017accurate}.
Comparing with the $GW$ approximation containing correlated electron-hole pairs
in the screened interaction, which are drawn as ring diagrams, the T-Matrix
approximation describes correlations of two electrons or  through Coulomb and
exchange interactions, which can be drawn as ladder diagrams. Thus the T-Matrix
approximation can be considered as the pp counterpart of $GW$. In the first
application of the T-Matrix approximation to molecular systems, the one-shot
calculation of the T-Matrix approximation (denoted as $G_0T_0$) was carried out\cite{zhang2017accurate}.
It was shown that the T-Matrix approximation has very good accuracy, with
much better agreements with experimental valence states than the bare KS
eigenvalues. However, as will be shown below, the QP equation of the $G_0T_0$
has multiple solutions in the core region, which gives erroneous CLBEs.
Beside the problematic behavior in computing CLBEs, the $G_0T_0$ method also
suffers from the strong dependence of the starting DFAs\cite{zhang2017accurate}.
For valence and core calculations, $G_0T_0$@HF predicts accurate results with a
mean absolute error (MAE) of $0.54$ \,{eV} but $G_0T_0$@PBE gives a relatively
large MAE of $1.90$ \,{eV}. Motivated by the success of using the RS Green's
function in $GW$\cite{jin2019renormalized}, in this work, we introduce the RS
Green's function to the T-Matrix method, and investigate valence and core state
properties of molecular systems. \par

We first review the self-energy from the $GW$ approximation and the T-Matrix
approximation. In the time domain, the real-space $GW$ self-energy is expressed
as the product of the Green's function and the two-point screened interaction
\begin{equation}\label{eq:gwse}
    \Sigma^{GW}(1,2) = iG(1,2)W(1,2)\text{,}
\end{equation}
where the variables $1$, $2$ are shorthand notations for combined
space-time-spin variables\cite{hedin1965new, martin2016interacting}. The
screened interaction $W$ can be formulated with eigenvalues and eigenvectors of
the phRPA matrix\cite{van2013gw}. \par

The T-Matrix approximation generalizes the two-point interaction $W(1,2)$ in
$GW$ to a four-point effective interaction $T(1,3;2,4)$, which
makes the T-Matrix self-energy an integral of the Green's function and $T$
\begin{equation}\label{eq:tmse}
    \Sigma^{\text{T}}(1,2) = i\int d 3 d 4 G(4,3)T(1,3;2,4)\text{.}
\end{equation}
In terms of diagrams, the T-Matrix approximation corresponds to an infinite
summation of ladder diagrams with corresponding exchange terms\cite{martin2016interacting,baym1961conservation, danielewicz1984quantum}.
Comparing to the infinite summation of ring diagrams in $GW$, which represents
screened interactions, the ladder diagrams indicate scattering interactions
between pairs of particles and between pairs of holes. The T-Matrix
approximation is exact up to the second order because of its proper description
of second-order exchange diagrams, which are missing in the $GW$
approximation. \par

The Fourier transform of the correlation part of the self-energy
$\Sigma^{\text{T}}$ in the frequency domain is given as\cite{zhang2017accurate}
\begin{align}\label{eq:tmsefreq}
    \begin{split}
       \Sigma^{\text{T}}_{\text{c}}(p,q,\omega)
       &= \sum_m \sum_i \frac{\langle pi | \chi^{N+2}_m\rangle \langle qi |
       \chi^{N+2}_m\rangle}{\omega + \epsilon_i - \omega^{N+2}_m + i\eta} \\
       & + \sum_m \sum_a \frac{\langle pa | \chi^{N-2}_m\rangle \langle qa |
       \chi^{N-2}_m\rangle}{\omega + \epsilon_a - \omega^{N-2}_m - i\eta}
       \text{.}
    \end{split}
\end{align}
Here, we use $i$, $j$, $k$, $l$ for occupied orbitals, $a$, $b$, $c$, $d$ for
virtual orbitals, $p$, $q$, $r$, $s$ for general orbitals and $m$ for the index
of the excitation from the ppRPA. The transition density in
Equation.\ref{eq:tmsefreq} is
\begin{align}\label{eq:transitiondensity}
    \langle pi | \chi^{N+2}_m\rangle =& \sum_{c<d}\langle pi||cd \rangle
    X^{N+2,m}_{cd} + \sum_{k<l}\langle pi||kl \rangle Y^{N+2,m}_{kl} \\
    \langle pa | \chi^{N-2}_m\rangle =& \sum_{c<d}\langle pa||cd \rangle
    X^{N-2,m}_{cd} + \sum_{k<l}\langle pa||kl \rangle Y^{N-2,m}_{kl} \text{,}
\end{align}
where $X^{N\pm 2}_m$, $Y^{N\pm 2}_m$ and $\omega^{N\pm 2}_m$ are two-electron
addition/removal eigenvectors and eigenvalues of the ppRPA matrix equation\cite{ring2004nuclear, van2013exchange, van2014exchange}
\begin{equation}\label{eq:pprpa}
    \begin{bmatrix}\mathbf{A} & \mathbf{B}\\
    \mathbf{B}^{\text{T}} & \mathbf{C}
    \end{bmatrix}\begin{bmatrix}\mathbf{X}\\
    \mathbf{Y}
    \end{bmatrix}=\omega^{N\pm2}\begin{bmatrix}\mathbf{I} & \mathbf{0}\\
    \mathbf{0} & \mathbf{-I}
    \end{bmatrix}\begin{bmatrix}\mathbf{X}\\
    \mathbf{Y}
    \end{bmatrix}\text{,}
\end{equation}
with
\begin{align}
    A_{ab,cd} & =\delta_{ac}\delta_{bd}(\epsilon_{a}+\epsilon_{b})+\langle
    ab||cd\rangle\text{,}\\
    B_{ab,kl} & =\langle ab||kl\rangle\text{,}\\
    C_{ij,kl} & =-\delta_{ik}\delta_{jl}(\epsilon_{i}+\epsilon_{j})+\langle
    ij||kl\rangle\text{.}
\end{align}
In above equations, the antisymmetrized two-electron integral
$\langle pq||rs\rangle$ is defined as
\begin{align}
   & \langle pq||rs\rangle = \langle pq|rs\rangle - \langle qp|rs\rangle \\
    & = \int d x_1 d x_2 \frac{\phi^*_p(x_1)\phi^*_q(x_2)(1 -
    \hat{P}_{12})\phi_r(x_1)\phi_s(x_2)}{|r_1 - r_2|}
\end{align}
where $x$ indicates both spatial and spin coordinates.
\par

In the $G_0T_0$ approach, the QP energies are calculated by the linearized QP equation\cite{zhang2017accurate}
\begin{equation}\label{eq:qpequation}
    \epsilon^{\text{QP}}_{p}=\epsilon_{p}^{\text{SCF}} + Z_p \langle
    p|\Sigma^{\text{T}}(p,p,\epsilon_{p}^{\text{SCF}})-v_{\text{xc}}|p\rangle
    \text{,}
\end{equation}
where the linearization factor $Z_p$ is
\begin{equation}
    Z_p =  \left(1- \left.\frac{\partial \Sigma^{\text{T}}(p,p,\omega)}{\partial
    \omega} \right|_{\omega = \epsilon_p^{\text{SCF}}}\right)^{-1} \text{.}
\end{equation}
Here $\{ \epsilon_{p}^{\text{SCF}} \}$ is a set of KS eigenvalues from the
self-consistent field (SCF) DFA calculation.

The one-shot $G_0T_0$ method has an undesired dependence on the choice of the
DFA, because the contributions of the singles, namely the effects of electron
exchange in $\Sigma^{\text{T}}$, are described only in a perturbative manner in
Eq.\ref{eq:qpequation}. The RS approach sums all the contributions of the singles
through a one-particle space diagonalization and thus reduces the dependence on
the starting DFA, and at the same time uses density matrices from commonly used
DFAs\cite{jin2019renormalized}. We now apply the RS Green's function in
the T-Matrix method. To take advantage of the form of the HF self-energy to
eliminate the dependence on the orbital energies of the DFA, the RS Green's
function is defined as the solution of the two projected equations in the
occupied orbital subspace and the virtual
orbital subspace\cite{jin2019renormalized}
\begin{equation}
        P(G_{\text{RS}}^{-1})P = P(G_{0}^{-1})P +
        P(\Sigma_{\text{Hx}}[G_{0}]-v_{\text{Hxc}})P \text{,}
\end{equation}
and
\begin{equation}
        Q(G_{\text{RS}}^{-1})Q = Q(G_{0}^{-1})Q +
        Q(\Sigma_{\text{Hx}}[G_{0}]-v_{\text{Hxc}})Q \text{,}
\end{equation}
where $P=\sum_{i}^{occ}|\psi_{i}\rangle\langle\psi_{i}|$ is the projection into
the occupied orbital space and $Q=I-P$ is the projection into the virtual
orbital space. $\Sigma_{\text{Hx}}[G_{0}]$ means that the HF self-energy
consisting of Hartree and exchange parts is constructed from the KS density
matrix. Equivalently, the RS Green's function is obtained by using the DFA
density matrix in the HF Hamiltonian, namely $H_{\text{HF}}[G_{0}]$, and
solving two projected HF equations in the occupied/virtual subspaces\cite{jin2019renormalized}
\begin{equation}
        P(H_{\text{HF}}[G_{0}])P|\Psi_{i}^{\text{RS}}\rangle =
        \varepsilon_{i}^{\text{RS}}P|\Psi_{i}^{\text{RS}}\rangle
        \text{,} \label{semip}
\end{equation}
and
\begin{equation}
        Q(H_{\text{HF}}[G_{0}])Q|\Psi_{a}^{\text{RS}}\rangle
        =\varepsilon_{a}^{\text{RS}}Q|\Psi_{a}^{\text{RS}}\rangle
        \text{.} \label{semiq}
\end{equation}
The resulting RS Green's function is diagonal in the occupied and virtual
subspaces\cite{jin2019renormalized}
\begin{equation}
        G_{nm}^{\text{RS}}(\omega) =
        \delta_{nm}\frac{1}{\omega-\varepsilon_{n}^{\text{RS}} +
        i\eta\text{sgn}(\varepsilon_{n}^{\text{RS}}-\mu)} \text{.}
\end{equation}
Here $\mu$ is the chemical potential and $\eta$ is the broadening parameter.
The RS Green's function has been implemented in the $GW$ calculations in the
QM4D package\cite{jin2019renormalized, qm4d}. \par

 We now develop the RS Green's function for the T-Matrix method. In the
 T-Matrix approximation, we apply the RS Green's function in two ways:
 $G_{\text{RS}}T_{\text{RS}}$ and $G_{\text{RS}}T_0$. The $G_{\text{RS}}T_0$
 approach uses the Green's function as a new starting point and the generalized
 effective interaction $T$ is calculated with the KS Green's function. During
 the evaluation of the self-energy, the KS orbitals are used for simplicity, as
 advocated in the original work of $G_{\text{RS}}W_0$\cite{jin2019renormalized}.
 The validity of using just the KS orbitals is further confirmed in present
 work. In the Supporting Information, it is shown that IPs and CLBEs from
 $G_{\text{RS}}T_0$ and $G_{\text{RS}}T_{\text{RS}}$ using KS orbitals and RS
 orbitals provide essentially the same results. The exchange part of the
 $G_{\text{RS}}T_0$ self-energy and the $G_{\text{RS}}T_{\text{RS}}$
 self-energy is the same as $G_0T_0$ and the correlation part of the
 $G_{\text{RS}}T_0$ self-energy is
\begin{align}
    \begin{split}
      \Sigma^{G_{\text{RS}}T_0}_{\text{c}}(p,q,\omega) &=
      \sum_m \sum_i \frac{\langle pi | \chi^{N+2}_m\rangle \langle qi |
      \chi^{N+2}_m\rangle}{\omega + \epsilon_i^{\text{RS}} -
      \omega^{N+2}_m - i\eta} \\
      & + \sum_m \sum_a \frac{\langle pa | \chi^{N-2}_m\rangle \langle qa |
      \chi^{N-2}_m\rangle}{\omega + \epsilon_a^{\text{RS}} - \omega^{N-2}_m +
      i\eta} \text{.}
    \end{split}
\end{align}
Here we follow the approximation in the $G_{\text{RS}}W_0$, where DFA orbitals
are used for simplicity. As can be seen above, KS eigenvalues in the
denominators are simply replaced by RS eigenvalues. The QP equation of
$G_{\text{RS}}T_0$
\begin{equation}\label{eq:qpequation_grst0}
    \epsilon^{\text{QP}}_{p}=\epsilon_{p}^{\text{SCF}}+ Z_p \langle p |
    \Sigma^{G_{\text{RS}}T_0}(p,p,\epsilon_{p}^{\text{RS}}) -
    v_{\text{xc}}|p\rangle \text{,}
\end{equation}
which can be linearized by the factor $Z_p = (1- \frac{\partial
\Sigma^{\text{c}}(p,p,\omega)}{\partial \omega}|_{\omega =
\epsilon_p^{\text{RS}}})^{-1}$. \par

The $G_{\text{RS}}T_{\text{RS}}$ approach further calculates the generalized
effective interaction $T$ with the RS Green's function, which means RS
eigenvalues are used in the ppRPA calculations. The KS orbitals are also used
for simplicity without the loss of accuracy (See Ref.\cite{jin2019renormalized}
and the Supporting Information). The self-energy of
$G_{\text{RS}}T_{\text{RS}}$ is thus
\begin{align}
    \begin{split}
        \Sigma^{G_{\text{RS}}T_{\text{RS}}}_{\text{c}}(p,q,\omega) & =
        \sum_m \sum_i \frac{\langle pi | \chi^{\text{RS},N+2}_m\rangle \langle
        qi | \chi^{\text{RS},N+2}_m\rangle}{\omega + \epsilon_i^{\text{RS}} -
        \omega^{\text{RS},N+2}_m - i\eta} \\
        & + \sum_m \sum_a \frac{\langle pa | \chi^{\text{RS},N-2}_m\rangle
        \langle qa | \chi^{\text{RS},N-2}_m\rangle}{\omega +
        \epsilon_a^{\text{RS}} - \omega^{\text{RS},N-2}_m + i\eta} \text{.}
    \end{split}
\end{align}
Thus the QP equation of $G_{\text{RS}}T_{\text{RS}}$ is
\begin{equation}\label{eq:qpequation_grstrs}
    \epsilon^{\text{QP}}_{p}=\epsilon_{p}^{\text{SCF}} + Z_p \langle p |
    \Sigma^{G_{\text{RS}}T_{\text{RS}}}(p,p,\epsilon_{p}^{\text{RS}}) -
    v_{\text{xc}}|p\rangle \text{,}
\end{equation}
which can be linearized by the factor $Z_p = (1- \frac{\partial
\Sigma^{\text{c}}(p,p,\omega)}{\partial \omega}|_{\omega =
\epsilon_p^{\text{RS}}})^{-1}$.
\par

We implemented $G_{\text{RS}}T_{\text{RS}}$ and $G_{\text{RS}}T_0$ methods in
QM4D quantum chemistry package\cite{qm4d} to calculate IPs and CLBEs of
molecular systems. As discussed in benchmarks of the GW100 set for $GW$ methods\cite{van2015gw,caruso2016benchmark},
EAs of many molecules in the GW100 set are negatives, where the experimental
values are not available. Therefore, we here focus on the discussion
of IPs. Results of EAs can be found in the Supporting Information.
We use Cartesian basis sets and uses the resolution of
identity\cite{eichkorn1995auxiliary,weigend2006accurate, ren2012resolution} (RI)
technique to compute two-electron integrals in the T-Matrix method. The
broadening parameter $\eta$ is set as $1.0\times10^{-3}$ \,{A.U.} for numerical
stability considerations. The convergence criteria is set as $1.0\times10^{-5}$
\,{A.U.} in the iterative procedure to solve the QP equation (Eqs.
\ref{eq:qpequation}, \ref{eq:qpequation_grst0} and \ref{eq:qpequation_grstrs}).
The convergence with respect to the broadening parameter has been assured and
test data can be found in supporting information. The T-Matrix methods are
tested with different noninteracting references, including HF and a variety of functionals,
such as the generalized gradient approximation (GGA) functional PBE\cite{perdew1996generalized},
and hybrid functionals B3LYP\cite{lee1988development,beck1993density} and PBE0\cite{adamo1999toward,ernzerhof1999assessment}.
We benchmarked IPs and EAs of molecules in the GW100 set with def2-TZVPP\cite{weigend2005balanced}
basis set and CLBEs of molecules in the CORE65 set with def2-TZVP\cite{weigend2005balanced}
basis set. We excluded 35 large molecules in the GW100 set and 8 large molecules
in the CORE65 set because of their high computational cost. Corresponding RI
fitting basis sets\cite{weigend1998ri} are used. All basis sets are taken from
Basis Set Exchange\cite{pritchard2019new,schuchardt2007basis,feller1996role}.
Experiment values of IPs and EAs are taken from Setten's work\cite{van2015gw}.
Experiment values of CLBEs are taken from Golze's work\cite{golze2019accurate}.
More details and results can be found in the Supporting Information. \par

\begin{figure*}
  \includegraphics[width=0.95\textwidth]{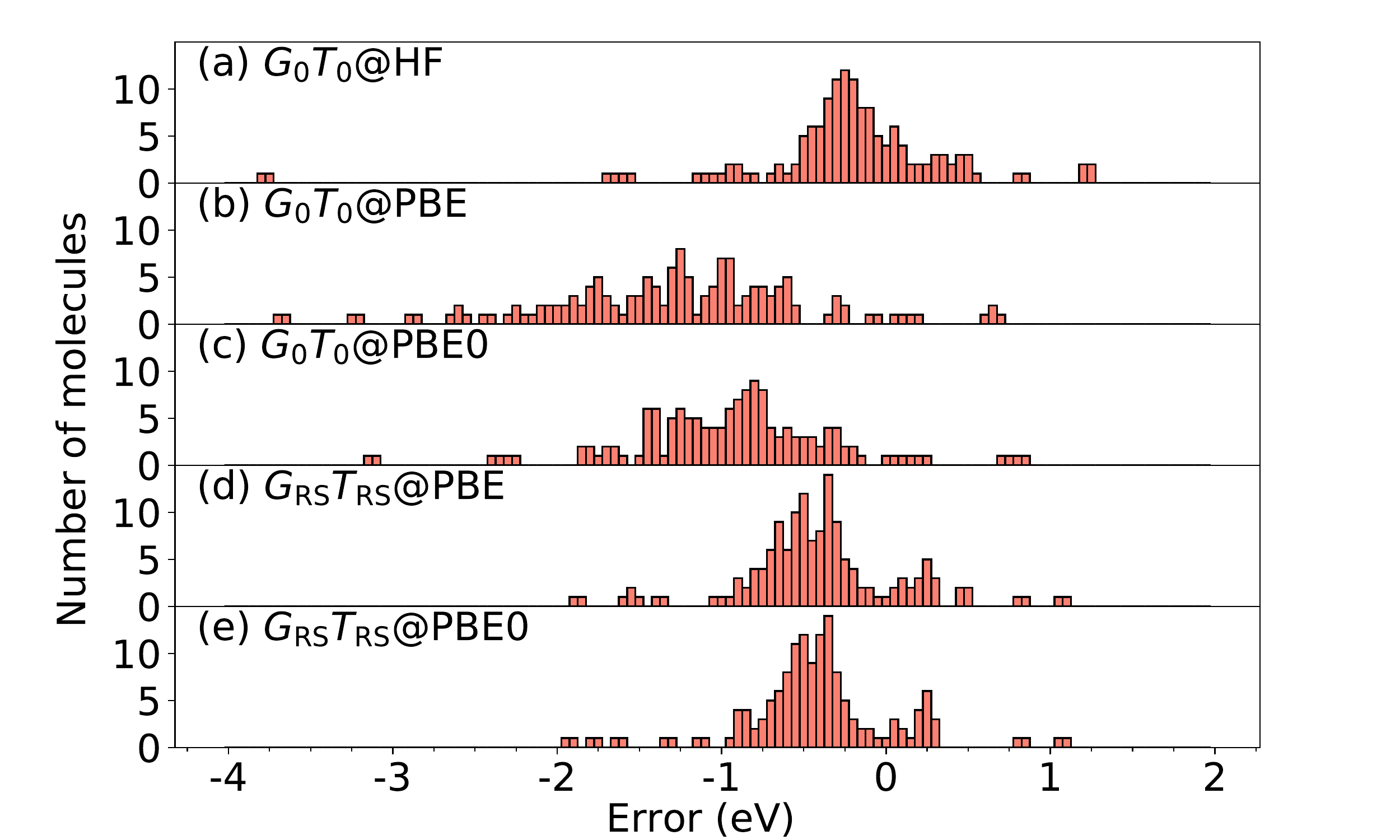}
  \caption{Distributions of errors with respect to the experimental IPs of the
  GW100 benchmark set from (a) $G_0T_0$@HF, (b) $G_0T_0$@PBE,
  (c) $G_0T_0$@PBE0, (d) $G_{\text{RS}}T_{\text{RS}}$@PBE,
  (e) $G_{\text{RS}}T_{\text{RS}}$@PBE0, where
  $\text{Error}_i = \text{IP}^{\text{theory}}_i - \text{IP}^{\text{exp}}_i$.}
  \label{fig:ip_error_distributaion}
\end{figure*}

We first examine $G_{\text{RS}}T_0$ and $G_{\text{RS}}T_{\text{RS}}$ methods
for molecular IP prediction. The IPs are obtained from $G_0T_0$,
$G_{\text{RS}}T_0$, $G_{\text{RS}}T_{\text{RS}}$ and $G_0W_0$ based on HF,
PBE and PBE0. The MAEs comparing to experiment
results can be seen in Table.\ref{gw100_ip_mae}. Our $G_0T_0$ results are
consistent with the previous work\cite{zhang2017accurate}, with $G_0T_0$@HF
having the smallest MAE. We also find that $G_0T_0$ with DFA references can
have improved accuracy with some fractions of the Hartree-Fock exchange.
As expected, both $G_{\text{RS}}T_0$@HF and $G_{\text{RS}}T_{\text{RS}}$@HF
show very similar results as $G_0T_0$@HF.
The MAEs of $G_{\text{RS}}T_0$ with both PBE and PBE0 are smaller comparing
with $G_0T_0$. The MAEs are reduced from $1.32$ \,{eV} and $0.98$ \,{eV} to
$1.03$ \,{eV} and $0.81$ \,{eV} when the RS Green's function is used as the new
starting point. The accuracy greatly improves when the generalized
effective interaction $T$ is also calculated with the RS Green's
function in the $G_{\text{RS}}T_{\text{RS}}$ approach. The MAEs of
$G_{\text{RS}}T_{\text{RS}}$ with PBE and PBE0 are reduced to $0.53$ \,{eV} and
$0.54$ \,{eV}.
In addition to the improved accuracy, the starting point dependence is greatly
reduced in the $G_{\text{RS}}T_{\text{RS}}$ approach. This can be seen in Fig.\ref{fig:ip_error_distributaion}.
For $G_0T_0$ results, different starting points show distinct error
distributions. The error distribution of  $G_0T_0$@HF is more centered around
zero than $G_0T_0$ combining with other DFAs. For $G_{\text{RS}}T_{\text{RS}}$,
using PBE or PBE0 as starting points provide very similar distributions, which
are also centered.
We also find that the distributions of $G_{\text{RS}}T_{\text{RS}}$
methods are in the range of $-2.0$ \,{eV} to $2.0$ \,{eV}. However, results of
$\text{As}_2$ and $\text{Br}_2$ molecules from $G_0T_0$@HF have
absolute errors of $4.11$ \,{eV} and $3.75$ \,{eV}, even though
$G_0T_0$@HF has a similar MAE to the $G_{\text{RS}}T_{\text{RS}}$ method.
Therefore, the $G_{\text{RS}}T_{\text{RS}}$ method has a better consistency
than $G_{0}T_{0}$@HF over the IP test sets. \par

\begin{table*}[t!]
\caption{Results of IPs in the GW100 set from $G_0T_0$, $G_{\text{RS}}T_0$,
$G_{\text{RS}}T_{\text{RS}}$ and $G_0W_0$ based on PBE,
PBE0 and HF (eV). "Max Error" means the largest absolute error in the GW100 set
predicted by the corresponding method.}
\label{gw100_ip_mae}\centering
\begin{tabular}{c ccc ccc ccc cc}
\hline
     & \multicolumn{3}{c} {$G_0T_0$} & \multicolumn{3}{c} {$G_{\text{RS}}T_0$} & \multicolumn{3}{c} {$G_{\text{RS}}T_{\text{RS}}$} & \multicolumn{2}{c} {$G_0W_0$} \\
  \cmidrule(l{0.5em}r{0.5em}){2-4} \cmidrule(l{0.5em}r{0.5em}){5-7} \cmidrule(l{0.5em}r{0.5em}){8-10} \cmidrule(l{0.5em}r{0.5em}){11-12}
           &  HF  &  PBE & PBE0 & HF   & PBE  & PBE0  & HF  & PBE  & PBE0 &  HF  & PBE \\
\hline
MAE        & 0.52 & 1.32 & 0.98 & 0.56 & 1.03 & 0.81 & 0.56 & 0.53 & 0.54 & 0.57 & 0.70\\
Max Error  & 4.11 & 3.66 & 3.08 & 4.11 & 3.01 & 2.67 & 4.11 & 1.91 & 1.84 & 3.70 & 2.93\\
\hline
\end{tabular}
\end{table*}

Next we apply $G_{\text{RS}}T_0$ and $G_{\text{RS}}T_{\text{RS}}$ methods to
core level calculations. We present CLBE results for molecules in the CORE65
set calculated with $G_0T_0$, $G_{\text{RS}}T_0$ and $G_{\text{RS}}T_{\text{RS}}$
combining with different starting points, including PBE, PBE0, B3LYP and HF.
The CORE65 set contains
$\text{C}_{1s}$, $\text{N}_{1s}$, $\text{O}_{1s}$ and $\text{F}_{1s}$
excitations from molecules that consist of up to 8 atoms. We excluded 8 large
molecules because of computational cost considerations. From Table.\ref{core65_mae}
it can been seen that using PBE as the starting point in $G_0T_0$ gives the
, largest error of around $15.0$ \,{eV}. The MAE is reduced when hybrid
functionals are used as the starting point. $G_0T_0$@PBE0 and $G_0T_0$@B3LYP
show similar MAEs around $9.0$ \,{eV}, which are still large. The
MAE of $G_0T_0$@HF is the smallest, which is similar to the conclusion for
valence calculations. The MAEs of $G_{\text{RS}}T_0$ are reduced from those of
$G_0T_0$ by about $2.0$ \,{eV}. It can be found that using the RS Green's
function in the T-Matrix further reduces errors for
all types of functionals, which are around $1.5$ \,{eV}. In addition, the
starting-point dependence is significantly reduced. All starting points give
MAEs smaller than $2.0$ \,{eV}. This shows that the
$G_{\text{RS}}T_{\text{RS}}$ method has the best accuracy and consistency
compared with other methods.

\begin{table}
  \caption{MAEs of CLBEs in the CORE65 set from $G_{0}T_{0}$, $G_{\text{RS}}T_0$, $G_{\text{RS}}T_{\text{RS}}$, $G_{0}W_{0}$ based on PBE, B3LYP, PBE0 and HF (eV).}
  \label{core65_mae}\centering
  \begin{tabular}{c|cccc}
    \hline
                                &PBE    & PBE0  &B3LYP & HF\\
    \hline
    $G_0T_0$                    & 14.97 & 7.80 & 9.34  & 3.74 \\
    $G_{\text{RS}}T_0$          & 12.21 & 6.47 & 7.55  & 3.74 \\
    $G_{\text{RS}}T_{\text{RS}}$& 1.53  & 2.06 & 1.66  & 3.74 \\
    $G_0W_0$                    &       & 5.06 & 5.96  & 5.67 \\
    \hline
  \end{tabular}

  \textsuperscript{\emph{a}} Cartesian def2-TZVP basis set is used.
  \textsuperscript{\emph{b}} Geometries and reference values are taken from
  Golze's work\cite{golze2020accurate}.
\end{table}

\begin{figure*}
  \includegraphics[width=0.95\textwidth]{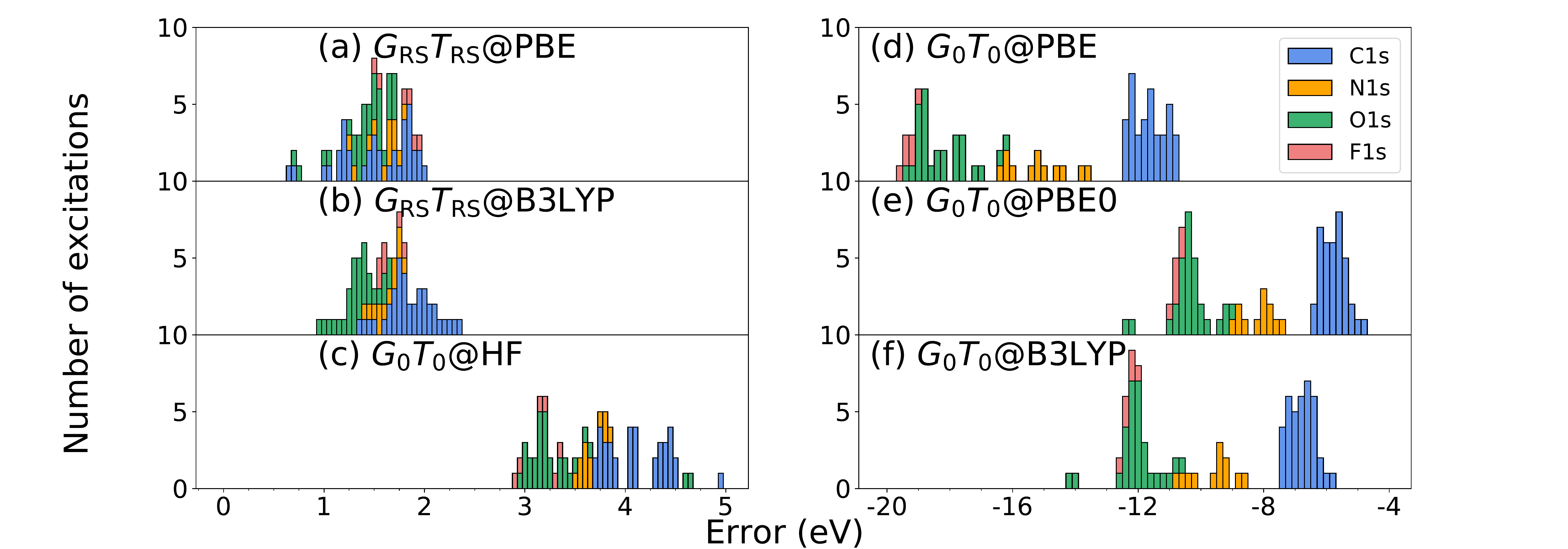}
  \caption{Distributions of errors with respect to the experimental CLBEs in
  the CORE65 benchmark set from (a) $G_{\text{RS}}T_{\text{RS}}$@PBE,
  (b) $G_{\text{RS}}T_{\text{RS}}$@B3LYP, (c) $G_0T_0$@HF, (d) $G_0T_0$@PBE,
  (e) $G_0T_0$@PBE0, (f) $G_0T_0$@B3LYP, where $\text{Error}_i =
  \text{CLBE}^{\text{theory}}_i - \text{CLBE}^{\text{exp}}_i$. The histograms
  are stacked.}
  \label{fig:core_error_distributaion}
\end{figure*}

The distributions of errors with respect to experimental values of CLBE from
$G_{\text{RS}}T_{\text{RS}}$ and $G_0T_0$ with different starting points are
shown in Fig.\ref{fig:core_error_distributaion}. It can be seen that the
distributions of $G_{\text{RS}}T_{\text{RS}}$ methods are the most centered.
The distributions of $G_0T_0$ methods are very different
when different starting points are used. Our calculations indicate that
$G_{\text{RS}}T_{\text{RS}}$ overestimate CLBEs only for less than $2.0$
\,{eV}, which clearly outperforms $G_0T_0$ (with $-20$ \,{eV} to $3$ \,{eV}
errors). $G_{\text{RS}}T_{\text{RS}}$ combining with different functionals
shows similar error distributions. This again indicates that functional
dependence is reduced significantly in the $G_{\text{RS}}T_{\text{RS}}$ method.

\begin{table*}
\caption{MAEs of relative CLBEs in the CORE65 set from $G_0T_0$ and $G_{\text{RS}}T_{\text{RS}}$ based on PBE, B3LYP, PBE0 and HF (eV). The relative CLBEs are the shifts with respect to a reference molecule, $\Delta\textnormal{CLBE}=\textnormal{CLBE}-\textnormal{CLBE}_{\textnormal{ref\_mol}}$. \ce{CH4}, \ce{NH3}, \ce{H2O} and \ce{CF4} have been used as reference molecules for $\text{C}_{1s}$, $\text{N}_{1s}$, $\text{O}_{1s}$ and $\text{F}_{1s}$ respectively.}
\label{core65_relative_mae}\centering
\begin{tabular}{ccccccccc}
\hline
  &\multicolumn{4}{c} {$G_0T_0$} & \multicolumn{4}{c} {$G_{\text{RS}}T_{\text{RS}}$}\\
  \cmidrule(l{0.5em}r{0.5em}){2-5} \cmidrule(l{0.5em}r{0.5em}){6-9}
    & PBE  & PBE0 & B3LYP & HF   & PBE  & PBE0 & B3LYP & HF\\
  \hline
  C & 0.85 & 0.42 & 0.41  & 0.33 & 0.39 & 0.29 & 0.46  & 0.33\\
  N & 1.40 & 0.78 & 0.88  & 0.09 & 0.19 & 0.11 & 0.12  & 0.09\\
  O & 2.29 & 1.11 & 1.47  & 0.22 & 0.29 & 0.17 & 0.21  & 0.22\\
  F & 0.22 & 0.18 & 0.11  & 0.13 & 0.22 & 0.09 & 0.07  & 0.13\\
  \hline
\end{tabular}
\end{table*}

\begin{figure}
  \includegraphics[width=0.9\textwidth]{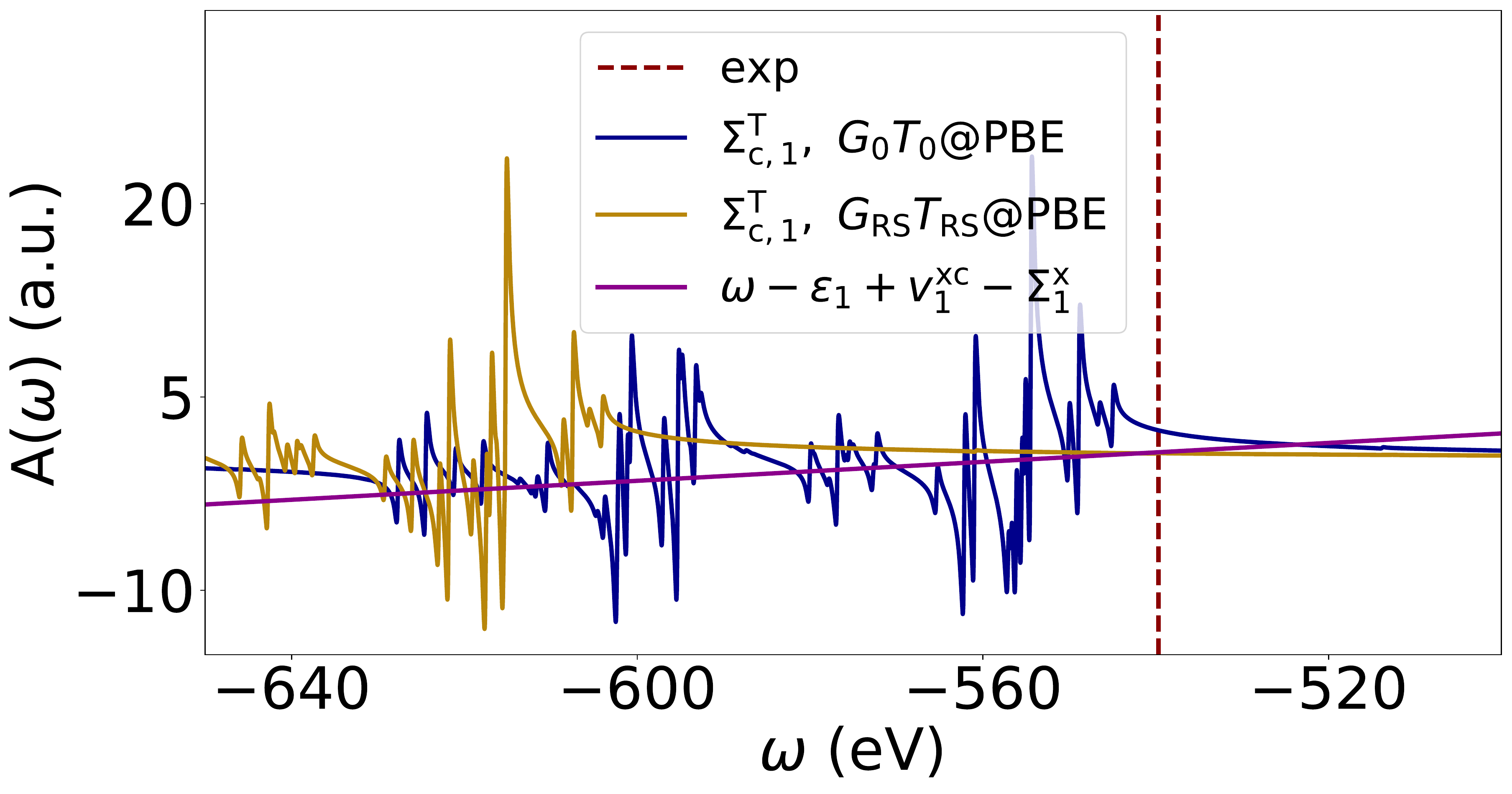}
  \caption{Graphical solutions of $\text{O}_{\text{1s}}$ excitation in the
  water molecule from $G_0T_0$@PBE and $G_{\text{RS}}T_{\text{RS}}$@PBE. The
  graphical solutions are found at intersections between
  $\omega-\epsilon_{\text{1s}} + v^{\text{xc}}_{\text{1s}} -
  \Sigma^{\text{x}}_{\text{1s}}$ (the green line) and the correlation part of
  $\text{O}_{\text{1s}}$ self-energy from $G_0T_0$@PBE (the blue line) or
  $G_{\text{RS}}T_{\text{RS}}$@PBE (the orange line). The vertical red line is
  the experimental value. def2-TZVP basis set is used.}
  \label{h2o_core_graphcial_compare}
\end{figure}

\begin{figure*}
  \includegraphics[width=0.9\textwidth]{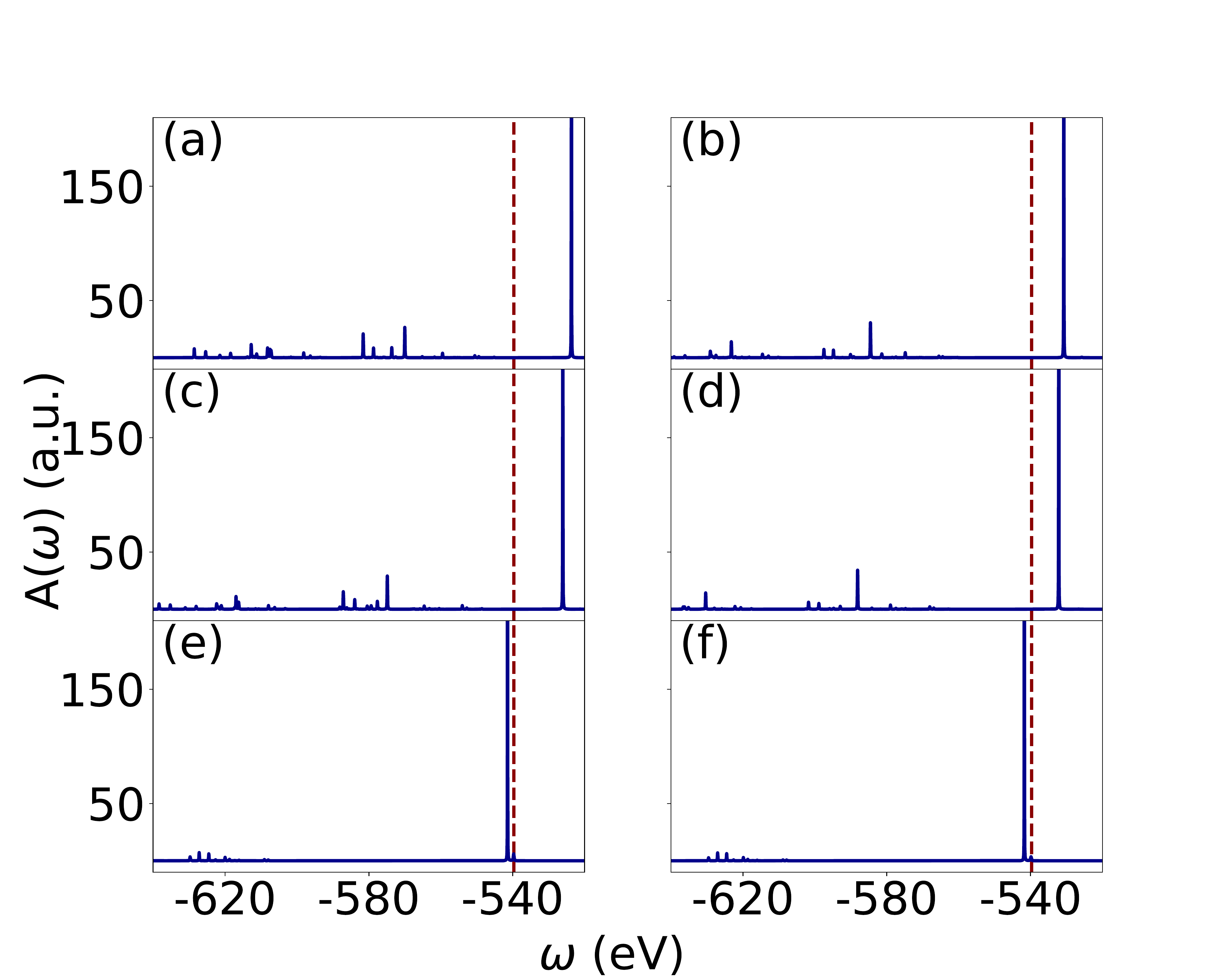}
  \caption{Spectral functions in the core region of the water molecule from (a)
  $G_0T_0$@PBE, (b) $G_0T_0$@PBE0, (c) $G_{\text{RS}}T_0$@PBE, (d)
  $G_{\text{RS}}T_0$@PBE0, (e) $G_{\text{RS}}T_{\text{RS}}$@PBE and (f)
  $G_{\text{RS}}T_{\text{RS}}$@PBE0. Blue lines are spectral functions and the
  red line is experimental value for the CLBE. def2-TZVP basis set is used.}
  \label{h2o_core_spectral}
\end{figure*}

The $G_{\text{RS}}T_{\text{RS}}$ method also provides improvement on predicting
relative CLBEs. As can be seen in Table.\ref{core65_relative_mae}, the MAEs of
relative CLBEs from $G_{\text{RS}}T_{\text{RS}}$ with all starting points are
below $0.50$ \,{eV}. For $\text{N}_{1s}$, $\text{O}_{1s}$ and
$\text{F}_{1s}$, $G_{\text{RS}}T_{\text{RS}}$ gives errors that are smaller
than $0.3$ \,{eV}. The starting point dependence in $G_0T_0$ is clearly shown.
$G_0T_0$@HF provides similar MAEs as $G_{\text{RS}}T_{\text{RS}}$, but $G_0T_0$
based on other KS starting points can give MAEs exceeding $2.00$ \,{eV}. The
dependence is greatly reduced in $G_{\text{RS}}T_{\text{RS}}$, as the
difference of MAEs between different starting points in
$G_{\text{RS}}T_{\text{RS}}$ are much smaller than those of $G_0T_0$. The good
performance of $G_{\text{RS}}T_{\text{RS}}$ is illustrated by the graphical
solutions of different approaches in Fig.\ref{h2o_core_graphcial_compare}.
In Fig.\ref{h2o_core_graphcial_compare}, the solutions of the QP equation is
found at intersections between the correlation part of $\text{O}_{\text{1s}}$
self-energy $\Sigma^{\text{T}}_{\text{c},1}$ and $\omega -
\epsilon_{\text{1s}}+v^{\text{xc}}_{\text{1s}} - \Sigma^{\text{x}}_{\text{1s}}$.
For $G_0T_0$@PBE, many intersections can be observed in the core
region. This means the QP state erroneously transfers spectral weight to
satellites, which leads to incorrect QP solutions. This error stems from
the underestimation of excitation energies in the ppRPA step at the PBE level.
When using RS Green's function, ppRPA excitation energies are
improved, thus satellites shift away from the correct QP states. Only one
intersection can be found in the core region, and it is our desired QP state.
The solutions from different approaches are also reflected in spectral
functions, as shown in Fig.\ref{h2o_core_spectral}. The equations for computing
spectral function in $G_{\text{RS}}T_0$ and $G_{\text{RS}}T_{\text{RS}}$ can be
found in the section.4 of the Supporting Information.
Spectral functions from $G_0T_0$@PBE,
$G_0T_0$@PBE0, $G_{\text{RS}}T_0$@PBE and $G_{\text{RS}}T_0$@PBE0 show multiple
peaks in the core region which correspond to multiple solutions in Fig.\ref{h2o_core_graphcial_compare}.
Spectral functions from $G_{\text{RS}}T_{\text{RS}}$@PBE and
$G_{\text{RS}}T_{\text{RS}}$@PBE0 show only one major peak, which corresponds
to the correct QP state. The erroneous behavior of multiple peaks in core
spectral functions was also found in $G_0W_0$ with commonly-used DFAs.
The correct behavior in $G_0W_0$ can be restored by tuning up the fraction of
exchange in the DFA\cite{golze2019accurate,golze2020accurate}.

In summary, we applied the RS Green's function in the T-Matrix method to
calculate valence and core states properties. Two methods were introduced: the
$G_{\text{RS}}T_0$ method that uses the RS Green's function as the reference
noninteracting Green's function and the $G_{\text{RS}}T_{\text{RS}}$ method
that further computes the generalized effective interaction with the the RS
Green's function.
$G_{\text{RS}}T_0$ and $G_{\text{RS}}T_{\text{RS}}$ methods were first examined
on valence state calculations by computing IPs in the GW100 set. It can be
found that the $G_{\text{RS}}T_{\text{RS}}$ method combining with PBE and PBE0
provides accurate results with MAEs of $0.53$ \,{eV} and $0.54$ \,{eV} which
are similar to that of $G_0T_0$@HF. However, $G_{\text{RS}}T_{\text{RS}}$ has a
much smaller error spread than $G_0T_0$@HF, and thus is a more reliable method
for predicting IPs.
$G_{\text{RS}}T_{\text{RS}}$ also greatly reduced the starting-point
dependence for IP calculations. The results on CLBEs of molecules in the CORE65
set show that the $G_{\text{RS}}T_{\text{RS}}$ method greatly outperforms
$G_0T_0$ and systematically eliminates the starting-point
dependence. The improvement of $G_{\text{RS}}T_{\text{RS}}$ comes from the fact
that the ppRPA excitation energies are larger, which ensures a unique solution
of the QP equation in the core region. This work demonstrates the capability of
the $G_{\text{RS}}T_{\text{RS}}$ method for predicting accurate QP energies for
molecular systems, both for valence and for core excitations.

ACKNOWLEDGMENTS: J. L. and Z.C. acknowledge the support from the National
Institute of General Medical Sciences of the National Institutes of Health
under award number R01-GM061870. W.Y. acknowledges the support from the
National Science Foundation (grant no. CHE-1900338).

\newpage

\begin{figure*}
  \includegraphics[width=0.95\textwidth]{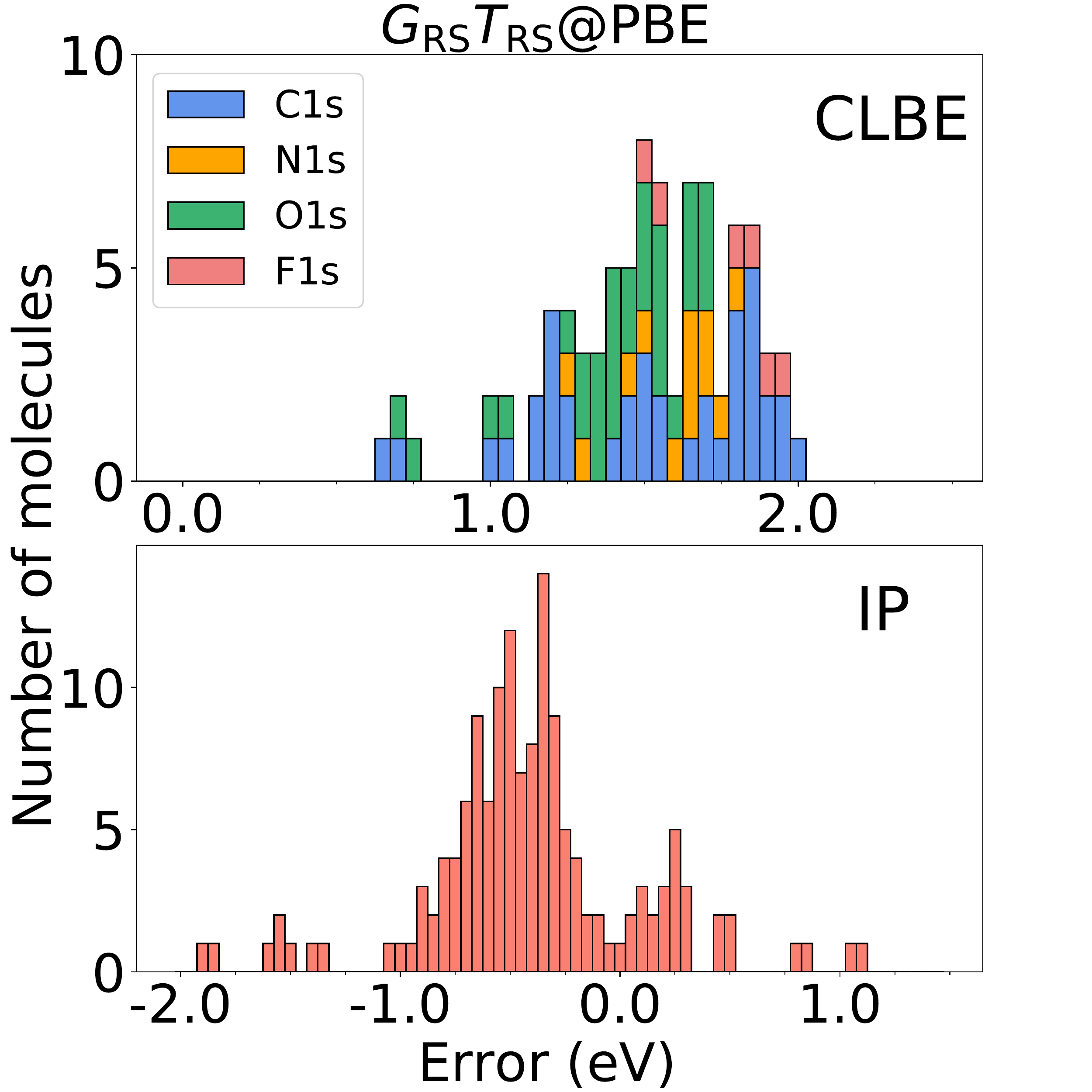}
  \caption{For Table of Contents Only”}
\end{figure*}

\newpage

\bibliography{ref.bib}

\end{document}


\section{Convergence of the Broadening Parameter}
\begin{figure}
    \centering
    \begin{subfigure}[h]{0.32\linewidth}
        \includegraphics[width=\linewidth]{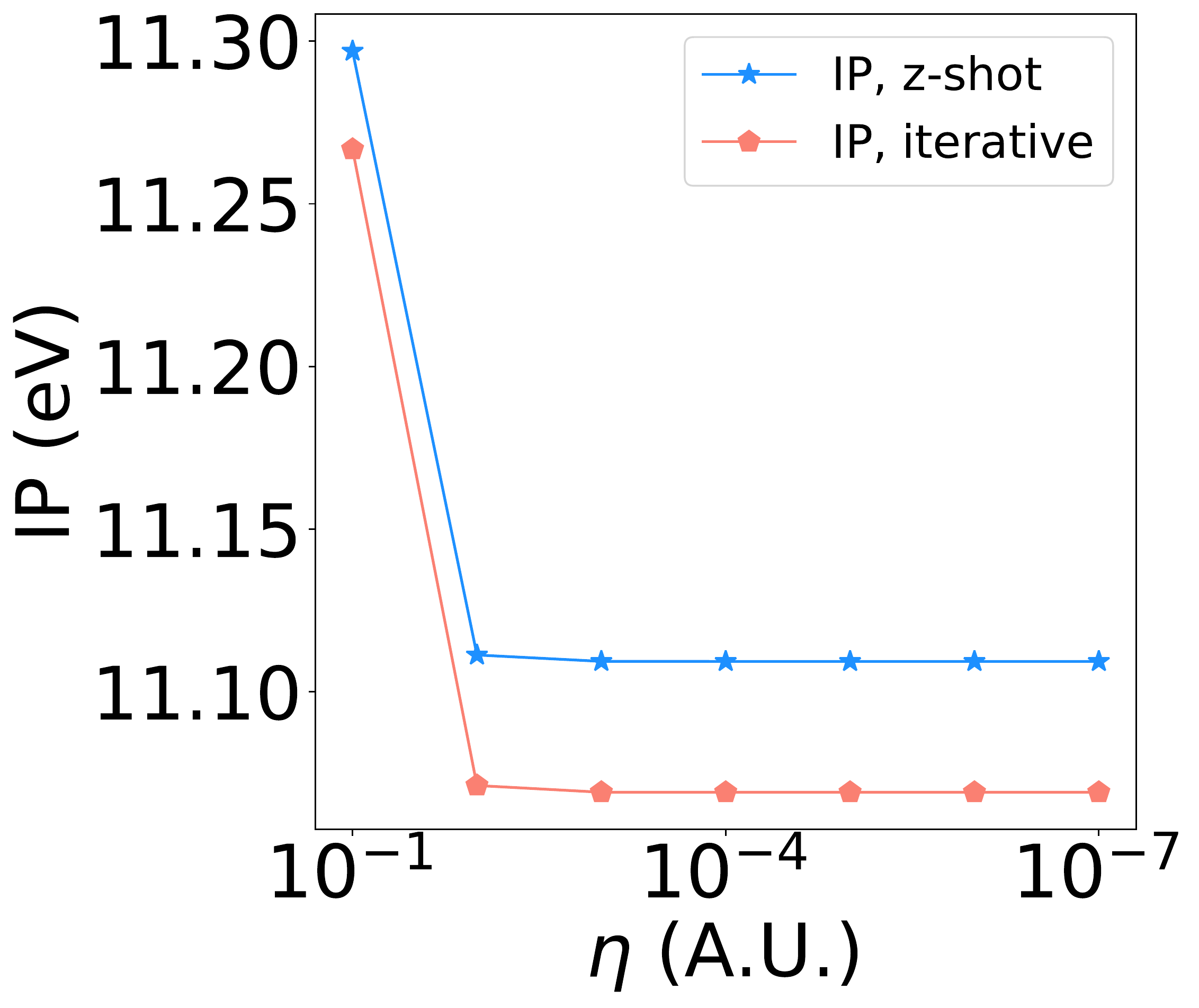}
        \caption{IP.}
        \label{eta_ip}
    \end{subfigure}
    \begin{subfigure}[h]{0.32\linewidth}
        \includegraphics[width=\linewidth]{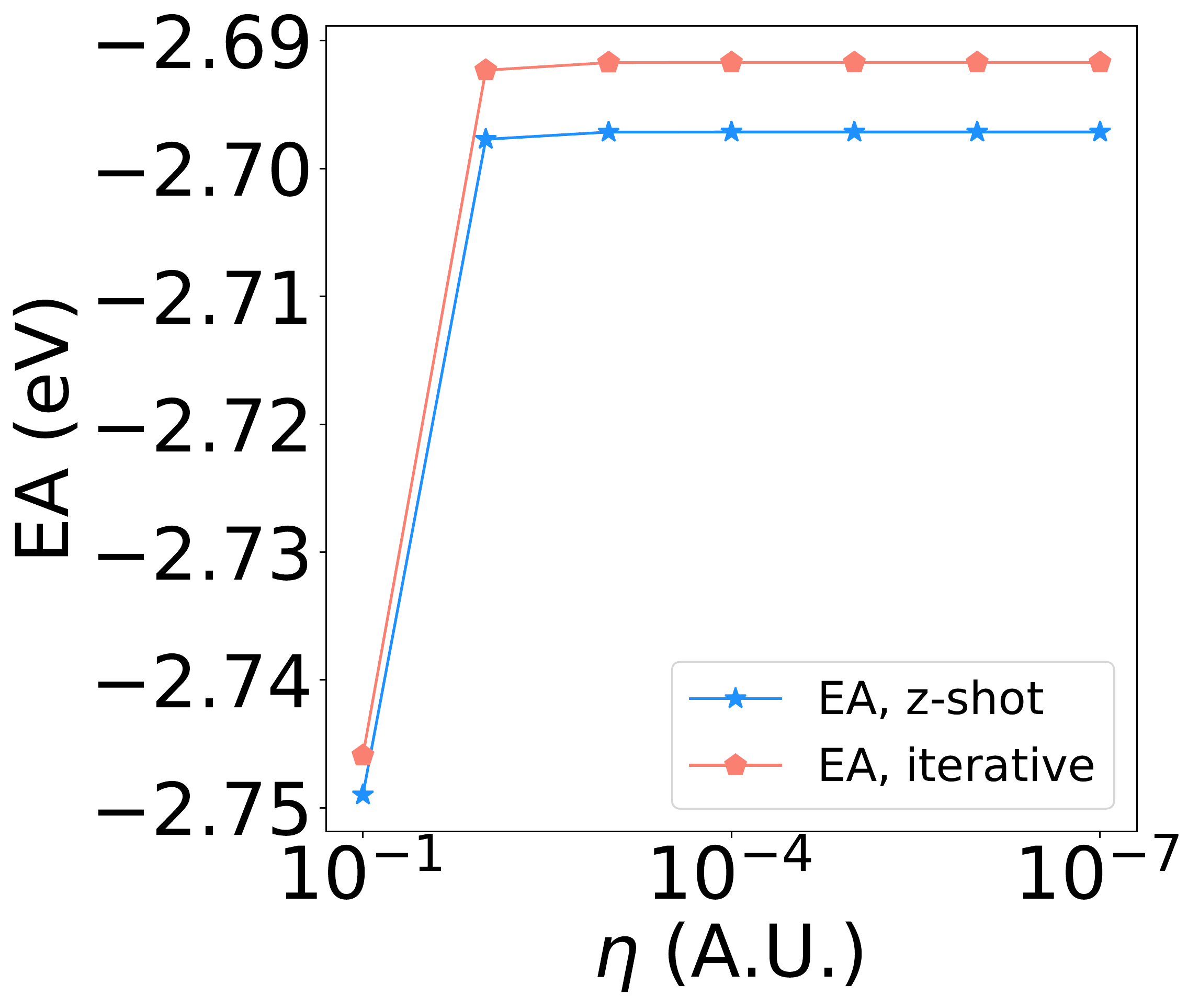}
        \caption{EA.}
        \label{eta_ea}
    \end{subfigure}
    \begin{subfigure}[h]{0.32\linewidth}
        \includegraphics[width=\linewidth]{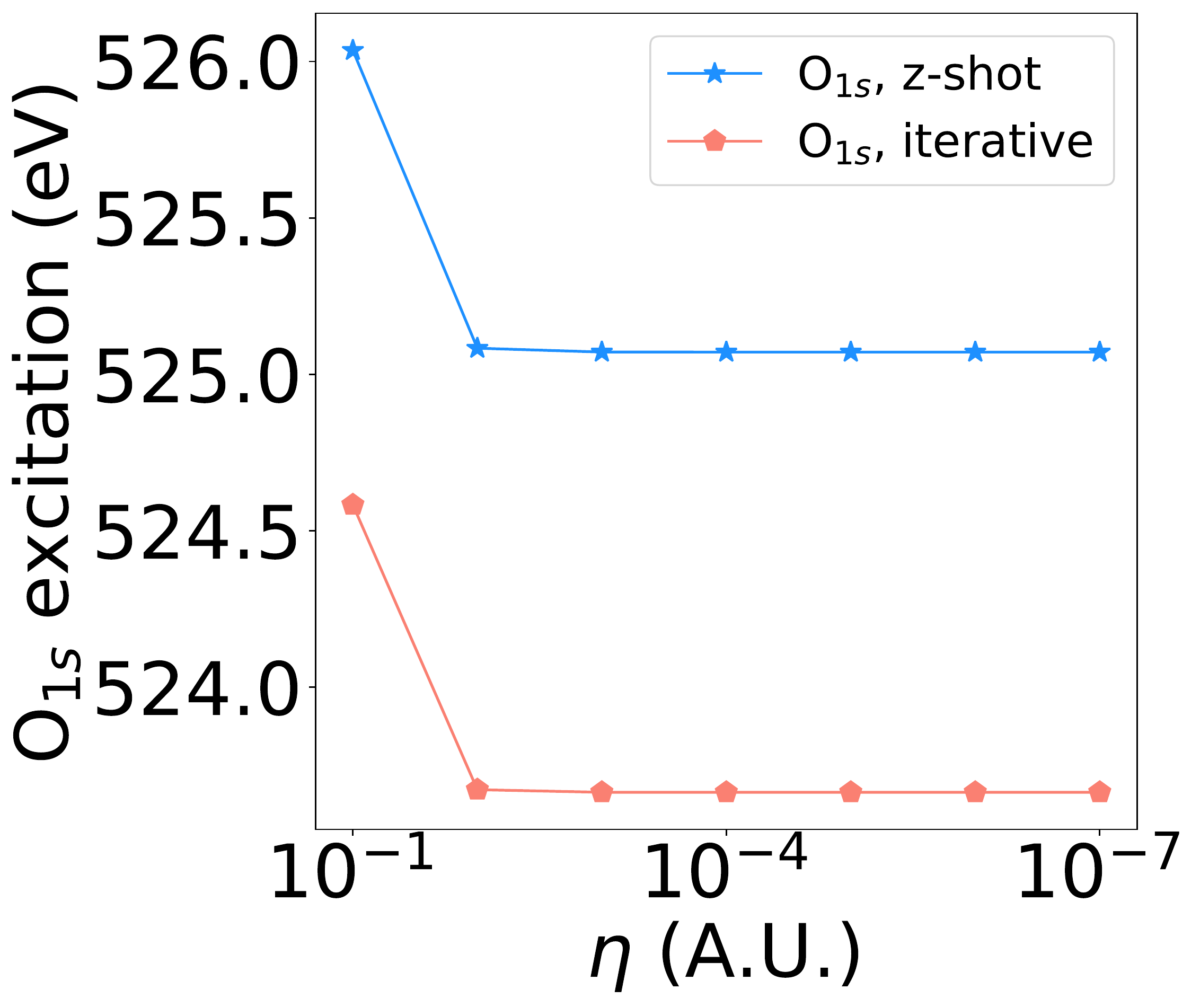}
        \caption{$\text{O}_{\text{1s}}$ CLBE.}
        \label{eta_core}
    \end{subfigure}
      \caption{IP (left), EA (middle) and $\text{O}_{\text{1s}}$ CLBE (right) of the water molecule with respect to the broadening parameter $\eta$, calculated by $G_0T_0$@PBE with def2-TZVPP basis set. The QP equation is solved by the z-shot method for the IP and the EA and by the iterative method for the CLBE.}
      \label{eta_convergence}
\end{figure}

The broadening parameter $\eta$ originated from the Fourier transform of the self-energy from the time-domain to the frequency domain affects solutions of the QP equations\cite{van2015gw, golze2019gw}. In principle $\eta$ is an infinitesimal positive number. The convergence of $\eta$ with respect to the ionization potential (IP), electron affinities (EA) and $\text{O}_{\text{1s}}$ core level binding energy (CLBE) of the water molecule calculated by $G_0T_0$@PBE with def2-TZVPP basis set are displayed in Fig.\ref{eta_convergence}.

As can be seen in Fig.\ref{eta_convergence},  when $\eta$ is smaller than $1.0\times10^{-2}$ A.U., differences drop below $1.0\times10^{-4}$ \,{eV}.

\section{Solutions of the Quasi-Particle Equation}
\begin{figure}
  \includegraphics[width=0.8\textwidth]{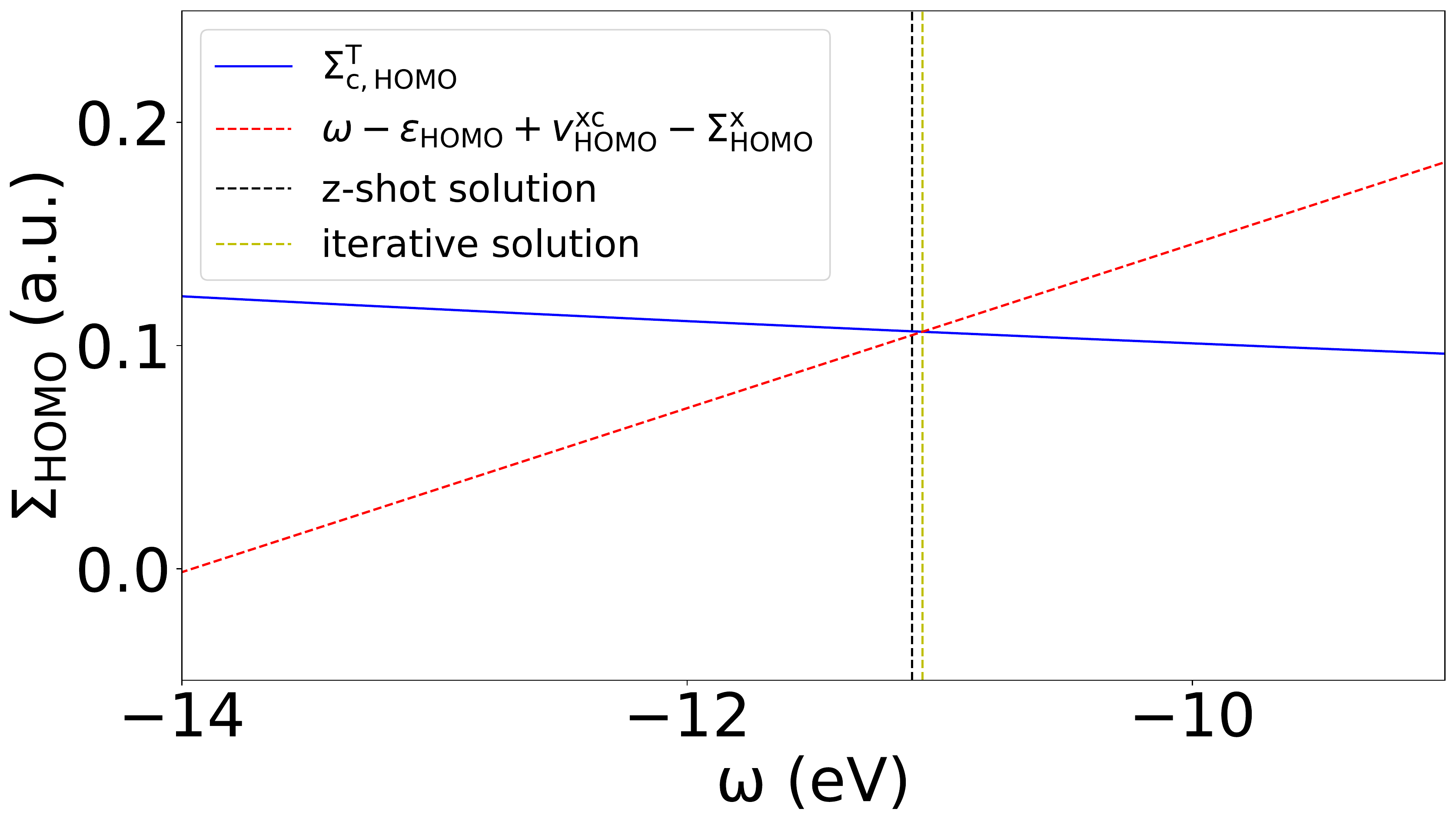}
  \caption{Solutions of HOMO from the z-shot method and the iterative method compared with the graphical solution, \ce{H2O} calculated by $G_0T_0$@PBE with def2-TZVPP basis set. The graphical solution is found at the intersection between the correlation part of HOMO self-energy $\Sigma^{\text{T}}_{\text{c,HOMO}}$ (the red line) and $\omega-\epsilon_{\text{HOMO}}+v^{\text{xc}}_{\text{HOMO}}-\Sigma^{\text{x}}_{\text{HOMO}}$ (the blue line) which is $-11.07$ eV. The black line stands for the solution from z-shot method which is $-11.11$ eV and the yellow line stands for the solution from iterative method which is $-11.07$ eV.}
  \label{homo_graphical}
\end{figure}

Practically, the z-shot method\cite{van2015gw, golze2019gw, martin2016interacting, zhang2017accurate} is used to solve the QP equation to avoid recalculation of the ppRPA matrix and the self-energy. This is valid when the difference between the z-shot method and the iterative method\cite{golze2019gw, martin2016interacting} of the QP equation is negligible. 
Here we discuss the error of using the z-shot method by investigating solutions of QP eigenvalues of the the water molecule from $G_0T_0$@PBE. The def2-TZVPP basis set is used. The solution of the HOMO from the z-shot method and the iterative method compared with the graphical solution are shown in Fig.\ref{homo_graphical}.
The graphical solution $-11.07$ eV is found at the intersection between the correlation part of the HOMO self-energy $\Sigma^{\text{T}}_{c,\text{HOMO}}$ (the red line) and $\omega-\epsilon_{\text{HOMO}}+v^{\text{xc}}_{\text{HOMO}}-\Sigma^{\text{x}}_{\text{HOMO}}$ (the blue line). As can be seen in Fig.\ref{homo_graphical}, the iterative method accurately reproduces the graphical solution at $-11.07$ eV. The z-shot method gives a solution at $-11.11$ eV which is slightly larger. The error from z-shot method $\Delta_{\text{z-shot}}$ is only $0.04$ eV. This means z-shot method is a good compromise between computational cost and accuracy for valence calculations.

Then we investigate solutions of $\text{O}_{\text{1s}}$ orbital from the z-shot method and the iterative method compared with the graphical solution
\begin{figure}
  \includegraphics[width=0.8\textwidth]{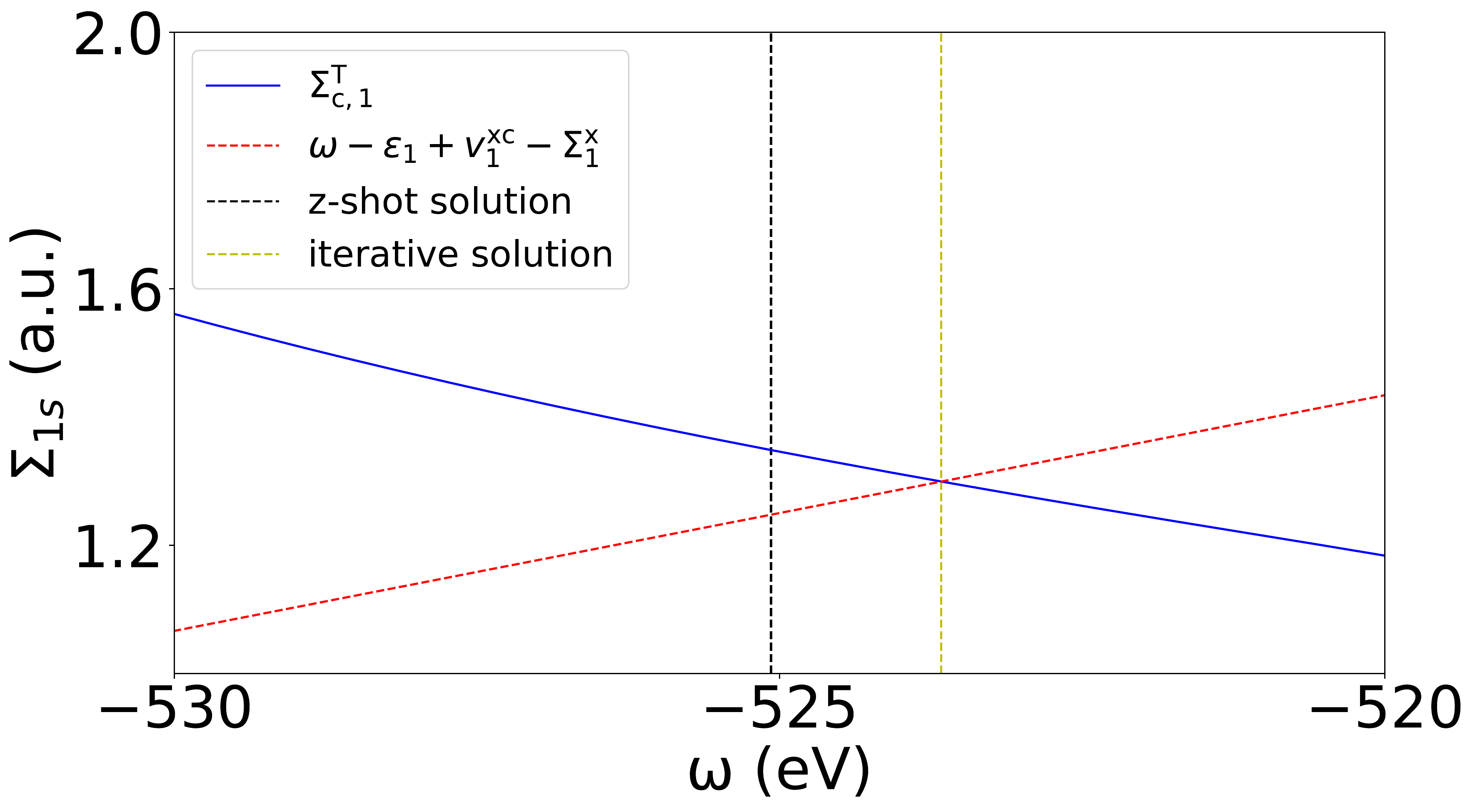}
  \caption{Solutions of $\text{O}_{\text{1s}}$ orbital from the z-shot method and the iterative method compared with the graphical solution, \ce{H2O} calculated by $G_0T_0$@PBE with def2-TZVPP basis set. The graphical solution is found at the intersection between the correlation part of $\text{O}_{\text{1s}}$ self-energy $\Sigma^{\text{T}}_{c,\text{1}}$ (the red line) and $\omega-\epsilon_{\text{1}}+v^{\text{xc}}_{\text{1}}-\Sigma^{\text{x}}_{\text{1}}$ (the blue line) which is $-523.67$ eV. The black line stands for the solution from z-shot method which is $-525.07$ eV and the yellow line stands for the solution from iterative method which is $-523.66$ eV.}
  \label{1s_graphical}
\end{figure}
The graphical solution is found at $-523.67$ eV. Again, the solution from the iterative method which is $-523.66$ eV agrees well with the graphical solution. However, the solution from the z-shot method which is $-525.07$ eV has a relatively large error of $1.40$ eV. The error is large because that the z-shot method from a Taylor expansion of the QP equation has larger errors for deeper states when KS eigenvalues is far from the corrected QP eigenvalues. 

\begin{figure}
  \includegraphics[width=0.6\textwidth]{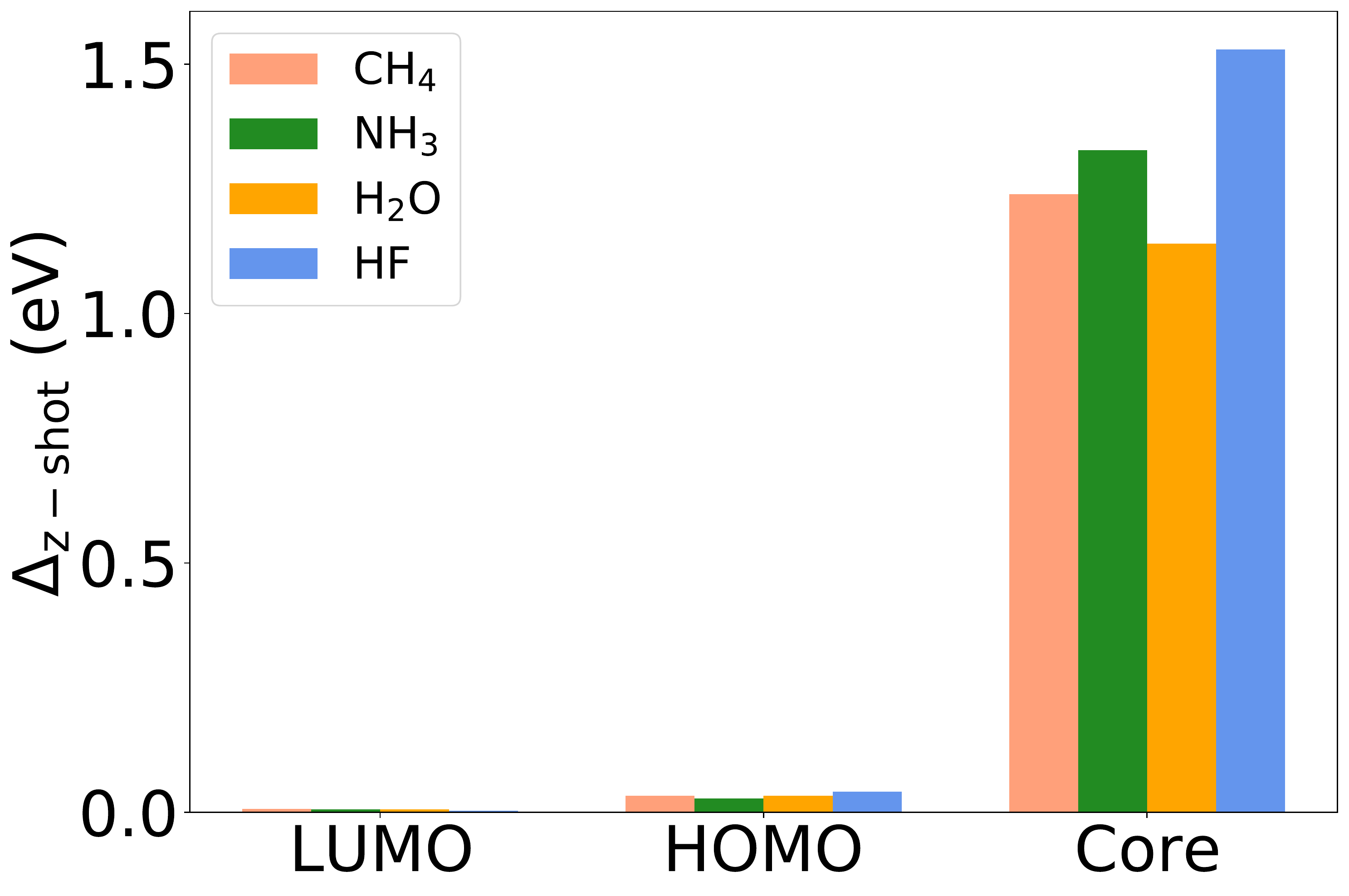}
  \caption{Errors from the z-shot method $\Delta_{\text{z-shot}}$ for LUMOs, HOMOs and Core states of different molecules calculated by $G_0T_0$@PBE with def2-TZVPP basis set. For \ce{CH4}, \ce{NH3}, \ce{H2O} and \ce{HF} molecule the core state is $\text{C}_{\text{1s}}$, $\text{N}_{\text{1s}}$, $\text{O}_{\text{1s}}$, $\text{F}_{\text{1s}}$ respectively.}
  \label{zshot_error_bar}
\end{figure}

We provide more examples of $\Delta_{\text{z-shot}}$ for valence and core calculation from three molecules in the Fig.\ref{zshot_error_bar}. The errors are similar to the errors of QP equation for $G_0W_0$\cite{golze2019gw}.
It is shown that for valence calculations, the z-shot method only has very small errors from $0.01$ eV to $0.03$ eV, but for deeper states calculations errors can be as large as $1.52$ eV. Thus in practical calculations, we use the z-shot method for valence calculations and iterative method for CLBE calculations.

\section{The Choice of Basis Set}
\begin{figure}
    \centering
    \begin{subfigure}[h]{0.3\linewidth}
        \includegraphics[width=\linewidth]{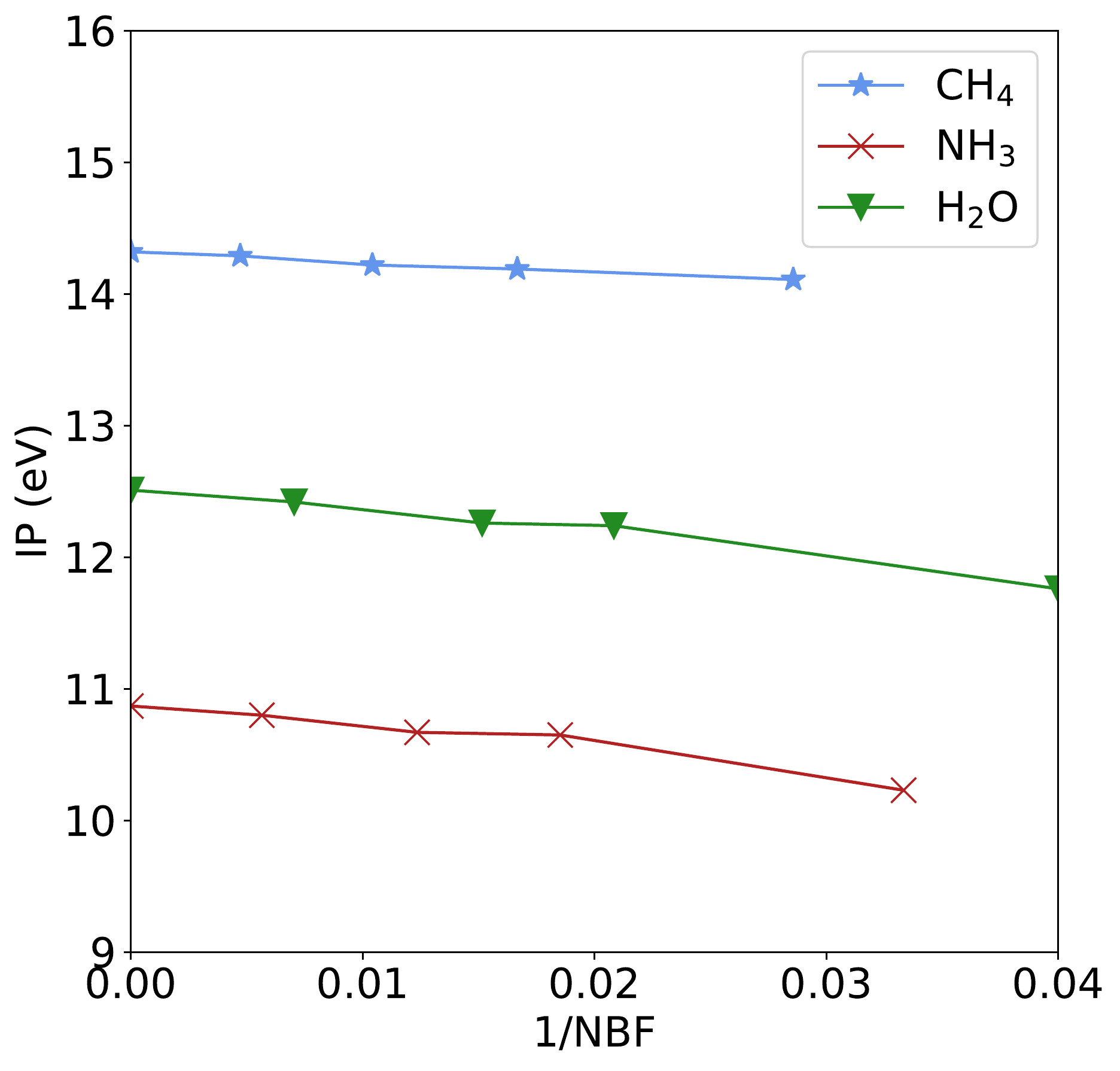}
        \caption{IP.}
        \label{basis_def_nbf_ip}
    \end{subfigure}
    \begin{subfigure}[h]{0.3\linewidth}
        \includegraphics[width=\linewidth]{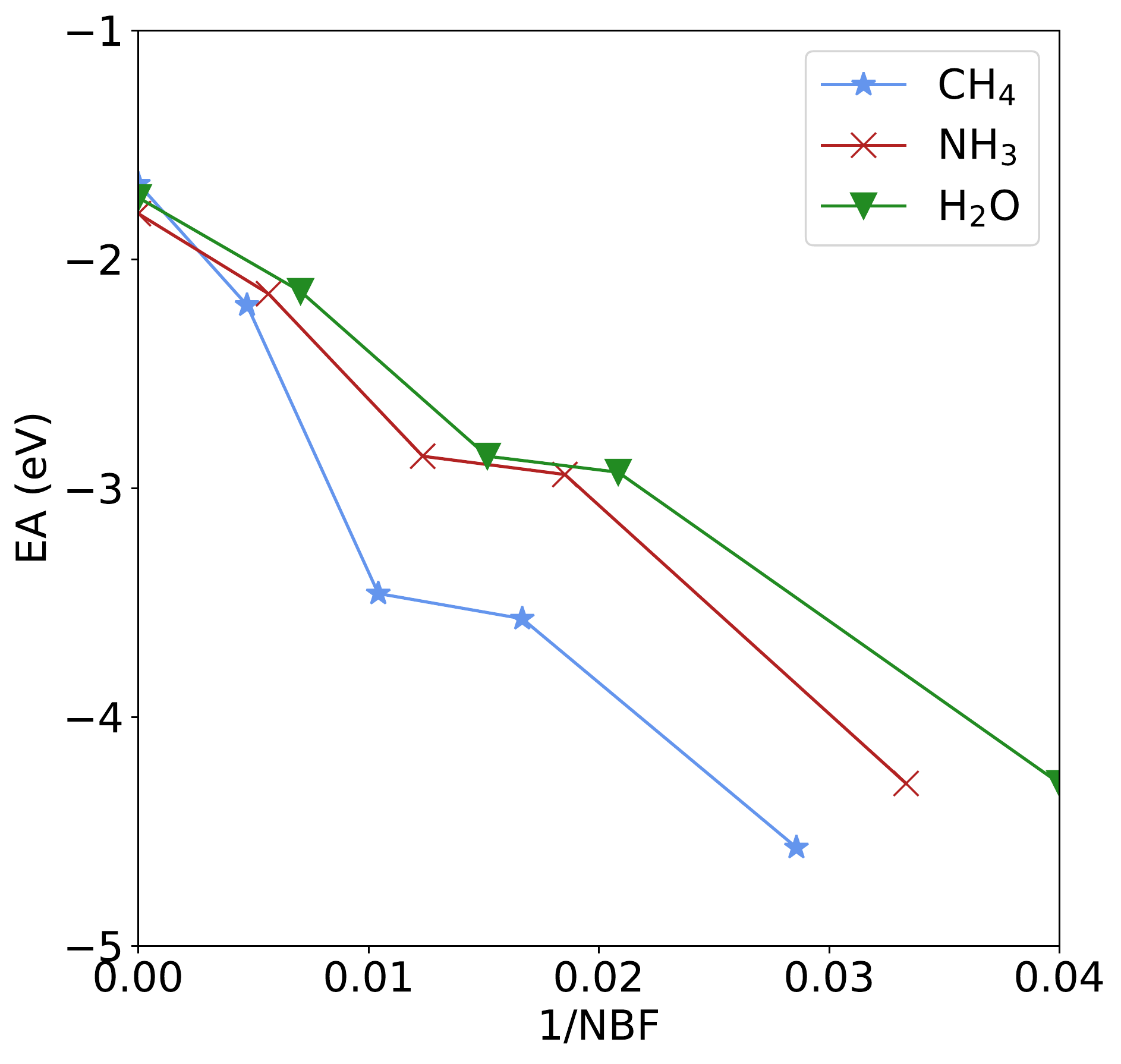}
        \caption{CLBEs.}
        \label{basis_def_nbf_ea}
    \end{subfigure}
    \begin{subfigure}[h]{0.3\linewidth}
        \includegraphics[width=\linewidth]{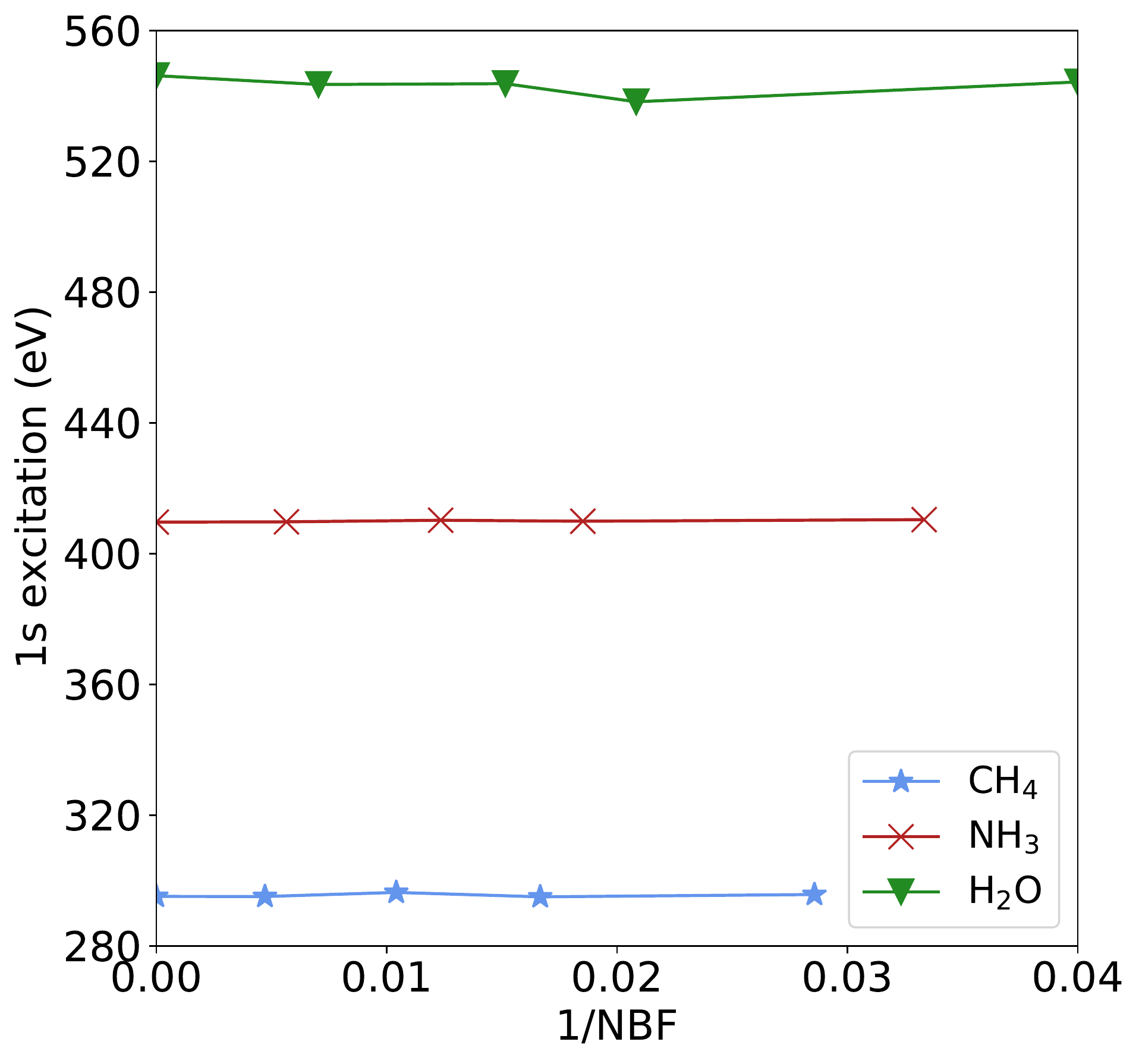}
        \caption{CLBEs.}
        \label{basis_def_nbf_core}
    \end{subfigure}
      \caption{IPs (left), EAs (middle) and CLBEs (right) of $\text{C}_{1s}$ in \ce{CH4}, $\text{N}_{1s}$ in \ce{NH3} and $\text{O}_{1s}$ in \ce{H2O} molecules (right) from $G_0T_0$@HF with def2 basis sets def2-SVP, def2-TZVP, def2-TZVPP and def2-QZVP with respect to the inverse of the number of basis functions. The complete basis set limit is obtained by a linear extrapolation of def2-TZVP and def2-QZVP results. The QP equation for IPs is solved by the z-shot method, for CLBEs is solved by the iterative method.}
\end{figure}

The convergence test for $G_0T_0$ with respect to basis sets is presented in this section. We compute valence and core states of 3 molecules with def2 basis sets. The IPs and CLBEs of \ce{CH4}, \ce{NH3} and \ce{H2O} molecules from $G_0T_0$@HF with def2 basis sets def2-SVP, def2-TZVP, def2-TZVPP and def2-QZVP and the complete basis set limit from a two-point extrapolation with respect to the inverse of the number of basis functions (NBF) are shown in the Fig.\ref{basis_def_nbf_ip}.
We find that results of basis set of triple zeta quality are reasonably converged. We choose def2-TZVPP basis set for valence calculation, which is also used in the benchmark of $GW$ methods\cite{caruso2016benchmark}. We choose def2-TZVP basis set for core level calculations. 

\FloatBarrier

\section{Spectral Function}\label{si:sec:spectralfunction}
The spectral function in the T-Matrix method is similar to that in $GW$\cite{golze2019gw, martin2016interacting}
\begin{equation}
    A(\omega)=\frac{1}{\pi}\sum_{n}\frac{|\text{Im}\Sigma_{n}(\omega)|}{[\omega-\varepsilon_{n}-(\text{Re}\Sigma_{n}(\omega)-v_{n}^{\text{xc}})]^{2}+[\text{Im}\Sigma_{n}(\omega)]^{2}} \text{.}
\end{equation}

In the $G_{\text{RS}}T_{0}$ and $G_{\text{RS}}T_{\text{RS}}$, because KS orbitals are used, the spectral function  has a similar form as that in $G_0T_0$
\begin{equation}
    A(\omega)=\frac{1}{\pi}\sum_{n}\frac{|\text{Im}\Sigma_{n}^{T}(\omega)|}{[\omega-\varepsilon_{n}^{\text{KS}}-(\text{Re}\Sigma_{n}^{T}(\omega)-v_{n}^{\text{xc}})]^{2}+[\text{Im}\Sigma_{n}^{T}(\omega)]^{2}}\text{.}
\end{equation}

\section{$G_{\text{RS}}T_0$ and $G_{\text{RS}}T_{\text{RS}}$ with KS and RS Orbitals}
\fontsize{9}{11}\selectfont{
\begin{longtable}{@{\extracolsep{\fill}}llccccc}
\caption{Ionization potentials of a subset of the GW100 set obtained with def2-TZVPP basis set, calculated by $G_{\text{RS}}T_0$@PBE and $G_{\text{RS}}T_{\text{RS}}$@PBE with KS and RS orbitals. All values in eV. }\\\toprule
 & & & \multicolumn{2}{c}{$G_{\text{RS}}T_0$}  & \multicolumn{2}{c}{$G_{\text{RS}}T_{\text{RS}}$} \\
 \cmidrule(l{0.5em}r{0.5em}){4-5} \cmidrule(l{0.5em}r{0.5em}){6-7}
name & formula & Exp & KS orbitals & RS orbitals & KS orbitals & RS orbitals\\
\hline 
\endfirsthead
\caption{Continued}\\\toprule
 & & & \multicolumn{2}{c}{$G_{\text{RS}}T_0$}  & \multicolumn{2}{c}{$G_{\text{RS}}T_{\text{RS}}$} \\
 \cmidrule(l{0.5em}r{0.5em}){4-5} \cmidrule(l{0.5em}r{0.5em}){6-7}
name & formula & Exp & KS orbitals & RS orbitals & KS orbitals & RS orbitals\\
\hline
\endhead
\bottomrule
\endlastfoot
helium                & He                         & 24.59 & 24.64 & 24.63 & 24.85 & 24.84 \\
neon                  & Ne                         & 21.56 & 20.46 & 20.51 & 21.46 & 21.49 \\
argon                 & Ar                         & 15.76 & 15.03 & 15.05 & 15.39 & 15.40 \\
krypton               & Kr                         & 14.00 & 13.39 & 13.44 & 13.68 & 13.72 \\
hydrogen              & $\text{H}_2$               & 15.43 & 16.16 & 16.15 & 16.27 & 16.26 \\
lithium dimer         & $\text{Li}_2$              & 4.73  & 4.98  & 4.96  & 5.01  & 5.00  \\
sodium dimer          & $\text{Na}_2$              & 4.89  & 4.65  & 4.64  & 4.67  & 4.66  \\
potassium dimer       & $\text{K}_2$               & 4.06  & 3.73  & 3.73  & 3.73  & 3.72  \\
phosphorus dimer      & $\text{P}_2$               & 10.62 & 9.79  & 9.79  & 9.95  & 9.93  \\
arsenic dimer         & $\text{As}_2$              & 10.00 & 9.03  & 9.05  & 9.15  & 9.16  \\
fluorine              & $\text{F}_2$               & 15.70 & 13.98 & 14.12 & 15.27 & 15.33 \\
chlorine              & $\text{Cl}_2$              & 11.49 & 10.53 & 10.56 & 10.97 & 10.98 \\
bromine               & $\text{Br}_2$              & 10.51 & 9.64  & 9.71  & 10.04 & 10.09 \\
methane               & C$\text{H}_4$              & 14.35 & 13.61 & 13.67 & 14.06 & 14.08 \\
ethyn                 & $\text{C}_2\text{H}_2$     & 11.49 & 10.59 & 10.60 & 10.89 & 10.88 \\
lithium hydride       & LiH                        & 7.90  & 8.04  & 8.05  & 8.22  & 8.22  \\
potassium hydride     & KH                         & 8.00  & 6.23  & 6.26  & 6.47  & 6.49  \\
borane                & B$\text{H}_3$              & 12.03 & 12.83 & 12.86 & 13.13 & 13.14 \\
ammonia               & N$\text{H}_3$              & 10.82 & 9.88  & 10.01 & 10.50 & 10.56 \\
phosphine             & P$\text{H}_3$              & 10.59 & 9.87  & 9.90  & 10.11 & 10.12 \\
arsine                & As$\text{H}_3$             & 10.58 & 9.70  & 9.74  & 9.92  & 9.95  \\
hydrogen sulfide      & S$\text{H}_2$              & 10.50 & 9.72  & 9.74  & 9.98  & 9.99  \\
hydrogen fluoride     & FH                         & 16.12 & 14.90 & 15.02 & 15.92 & 15.97 \\
hydrogen chloride     & ClH                        & 12.79 & 12.02 & 12.05 & 12.35 & 12.36 \\
lithium fluoride      & LiF                        & 11.30 & 10.27 & 10.45 & 11.47 & 11.54 \\
boron monofluoride    & BF                         & 11.00 & 10.35 & 10.38 & 10.58 & 10.59 \\
boron nitride         & BN                         &       & 10.63 & 10.62 & 11.18 & 11.14 \\
hydrogen cyanide      & NCH                        & 13.61 & 12.96 & 12.94 & 13.30 & 13.26 \\
hydrogen peroxide     & HOOH                       & 11.70 & 10.01 & 10.22 & 11.05 & 11.15 \\
water                 & $\text{H}_2$O              & 12.62 & 11.48 & 11.61 & 12.32 & 12.38 \\
carbon dioxide        & C$\text{O}_2$              & 13.77 & 12.41 & 12.52 & 13.24 & 13.26 \\
carbon monoxide       & CO                         & 14.01 & 13.04 & 13.14 & 13.67 & 13.70 \\
beryllium monoxide    & BeO                        & 10.10 & 8.77  & 8.93  & 9.77  & 9.82 
\end{longtable}
}

\fontsize{9}{11}\selectfont{
\begin{longtable}{@{\extracolsep{\fill}}llccccc}
\caption{Core level binding energies of a subset of the CORE65\cite{golze2019accurate} set obtained with def2-TZVPP basis set calculated by $G_{\text{RS}}T_0$@PBE and $G_{\text{RS}}T_{\text{RS}}$@PBE with KS and RS orbitals. All values in eV. }\\\toprule
 & & & \multicolumn{2}{c}{$G_{\text{RS}}T_0$}  & \multicolumn{2}{c}{$G_{\text{RS}}T_{\text{RS}}$} \\
 \cmidrule(l{0.5em}r{0.5em}){4-5} \cmidrule(l{0.5em}r{0.5em}){6-7}
name & formula & Exp & KS orbitals & RS orbitals & KS orbitals & RS orbitals\\
\hline 
\endfirsthead
\caption{Continued}\\\toprule
 & & & \multicolumn{2}{c}{$G_{\text{RS}}T_0$}  & \multicolumn{2}{c}{$G_{\text{RS}}T_{\text{RS}}$} \\
 \cmidrule(l{0.5em}r{0.5em}){4-5} \cmidrule(l{0.5em}r{0.5em}){6-7}
name & formula & Exp & KS orbitals & RS orbitals & KS orbitals & RS orbitals\\
\hline
\endhead
\bottomrule
\endlastfoot
C$\text{H}_4$                 & C1s                  & 290.844 & 282.19 & 283.47 & 292.07 & 293.39 \\
$\text{C}_2\text{H}_6$        & C1s                  & 290.714 & 281.58 & 282.91 & 292.39 & 292.89 \\
$\text{C}_2\text{H}_4$        & C1s                  & 290.823 & 281.33 & 282.58 & 292.69 & 293.19 \\
$\text{C}_2\text{H}_2$        & C1s                  & 291.249 & 281.14 & 282.74 & 293.01 & 293.56 \\
CO                            & O1s                  & 542.1   & 527.49 & 528.80 & 543.65 & 544.30 \\
CO                            & C1s                  & 296.229 & 288.05 & 289.24 & 298.14 & 298.54 \\
C$\text{O}_2$                 & O1s                  & 541.32  & 525.02 & 526.44 & 542.36 & 543.01 \\
C$\text{O}_2$                 & C1s                  & 297.699 & 288.61 & 289.77 & 298.75 & 299.22 \\
C$\text{H}_3$OH               & O1s                  & 538.88  & 524.11 & 525.58 & 540.43 & 541.21 \\
C$\text{H}_3$OH               & C1s                  & 292.3   & 283.55 & 284.69 & 294.26 & 294.71 \\
C$\text{H}_2$O                & O1s                  & 539.33  & 524.14 & 525.43 & 540.87 & 541.56 \\
C$\text{H}_2$O                & C1s                  & 294.38  & 285.82 & 286.93 & 296.40 & 296.83 \\
HCOOH                         & O1s\_OH              & 540.69  & 525.21 & 526.61 & 542.07 & 542.75 \\
HCOOH                         & O1s\_C=O             & 539.02  & 523.16 & 524.49 & 540.30 & 540.95 \\
HCOOH                         & C1s                  & 295.75  & 286.54 & 287.64 & 297.19 & 297.62 \\
$\text{H}_2$O                 & O1s                  & 539.7   & 526.00 & 527.35 & 541.42 & 541.99 \\
$\text{O}_3$                  & O1s\_midd            & 546.44  & 530.76 & 531.70 & 547.90 & 548.39 \\
$\text{O}_3$                  & O1s\_term            & 541.75  & 523.32 & 523.78 & 542.50 & 542.93 \\
$\text{N}_2$                  & N1s                  & 409.93  & 397.43 & 399.07 & 411.21 & 411.74 \\
N$\text{H}_3$                 & N1s                  & 405.52  & 394.09 & 395.44 & 407.30 & 407.89 \\
HCN                           & N1s                  & 406.8   & 393.63 & 394.96 & 408.28 & 408.89 \\
HCN                           & C1s                  & 293.5   & 283.73 & 285.69 & 295.04 & 295.53 \\
C$\text{H}_3$CN               & N1s                  & 405.58  & 392.06 & 393.54 & 407.24 & 407.87 \\
C$\text{H}_3$CN               & C1s\_C$\text{H}_3$   & 292.88  & 282.96 & 284.42 & 294.38 & 294.90 \\
C$\text{H}_3$CN               & C1s\_CN              & 292.60  & 282.96 & 284.40 & 294.17 & 294.70 \\
C$\text{H}_3\text{NH}_2$      & N1s                  & 405.17  & 393.10 & 394.59 & 406.85 & 407.49
\end{longtable}
}

\section{GW100 set results}
The IPs and EAs of molecules in the GW100\cite{van2015gw} set calculated from $G_{0}T_{0}$, $G_{\text{RS}}T_0$ and $G_{\text{RS}}T_{\text{RS}}$ based on PBE, PBE0 and HF with def2-TZVPP are presented in this section. Reference values are taken from Setten's work\cite{van2015gw}. 35 molecules are excluded considering computational cost. All calculations are performed by our software QM4D\cite{qm4d}.

\fontsize{9}{11}\selectfont{
\begin{longtable}{@{\extracolsep{\fill}}llcccccccccc}
\caption{Ionization potentials of the GW100 set\cite{van2015gw} obtained with def2-TZVPP basis set from $G_{0}T_{0}$, $G_{\text{RS}}T_0$ and $G_{\text{RS}}T_{\text{RS}}$ with different starting points including HF, PBE and PBE0. All values in eV. }\\\toprule
 & & & \multicolumn{3}{c}{$G_{0}T_{0}$} & \multicolumn{3}{c}{$G_{\text{RS}}T_0$}  & \multicolumn{3}{c}{$G_{\text{RS}}T_{\text{RS}}$} \\
 \cmidrule(l{0.5em}r{0.5em}){4-6} \cmidrule(l{0.5em}r{0.5em}){7-9} \cmidrule(l{0.5em}r{0.5em}){10-12}
name & formula & Exp & HF & PBE & PBE0 & HF & PBE & PBE0 & HF & PBE & PBE0  \\
\hline 
\endfirsthead
\caption{Continued}\\\toprule
 & & & \multicolumn{3}{c}{$G_{0}T_{0}$} & \multicolumn{3}{c}{$G_{\text{RS}}T_0$}  & \multicolumn{3}{c}{$G_{\text{RS}}T_{\text{RS}}$} \\
 \cmidrule(l{0.5em}r{0.5em}){4-6} \cmidrule(l{0.5em}r{0.5em}){7-9} \cmidrule(l{0.5em}r{0.5em}){10-12}
name & formula & Exp & HF & PBE & PBE0 & HF & PBE & PBE0 & HF & PBE & PBE0  \\
\hline
\endhead
\bottomrule
\endlastfoot
helium                & He                         & 24.59 & 24.75   & 24.55    & 24.63     & 24.75    & 24.64     & 24.68      & 24.75     & 24.85      & 24.82       \\
neon                  & Ne                         & 21.56 & 21.06   & 20.06    & 20.39     & 21.06    & 20.46     & 20.64      & 21.06     & 21.46      & 21.29       \\
argon                 & Ar                         & 15.76 & 15.53   & 14.83    & 15.02     & 15.53    & 15.03     & 15.15      & 15.53     & 15.39      & 15.36       \\
krypton               & Kr                         & 14.00 & 13.79   & 13.24    & 13.37     & 13.79    & 13.39     & 13.47      & 13.79     & 13.68      & 13.64       \\
hydrogen              & $\text{H}_2$               & 15.43 & 16.26   & 16.13    & 16.17     & 16.26    & 16.16     & 16.19      & 16.26     & 16.27      & 16.26       \\
lithium dimer         & $\text{Li}_2$              & 4.73  & 5.04    & 4.94     & 4.96      & 5.04     & 4.98      & 4.98       & 5.04      & 5.01       & 5.00        \\
sodium dimer          & $\text{Na}_2$              & 4.89  & 4.70    & 4.65     & 4.65      & 4.70     & 4.65      & 4.65       & 4.70      & 4.67       & 4.65        \\
sodium tetramer       & $\text{Na}_4$              & 4.27  & 3.93    & 3.70     & 3.76      & 3.93     & 3.74      & 3.79       & 3.93      & 3.78       & 3.80        \\
potassium dimer       & $\text{K}_2$               & 4.06  & 3.82    & 3.74     & 3.74      & 3.82     & 3.73      & 3.74       & 3.82      & 3.73       & 3.72        \\
nitrogen              & $\text{N}_2$               & 15.58 & 16.83   & 13.82    & 14.41     & 16.83    & 14.36     & 14.75      & 16.83     & 15.12      & 15.22       \\
phosphorus dimer      & $\text{P}_2$               & 10.62 & 10.20   & 9.69     & 9.83      & 10.20    & 9.79      & 9.90       & 10.20     & 9.95       & 9.96        \\
arsenic dimer         & $\text{As}_2$              & 10.00 & 5.89    & 8.96     & 9.07      & 5.89     & 9.03      & 9.12       & 5.89      & 9.15       & 9.16        \\
fluorine              & $\text{F}_2$               & 15.70 & 15.40   & 13.13    & 13.90     & 15.40    & 13.98     & 14.42      & 15.40     & 15.27      & 15.24       \\
chlorine              & $\text{Cl}_2$              & 11.49 & 11.28   & 10.17    & 10.51     & 11.28    & 10.53     & 10.74      & 11.28     & 10.97      & 10.99       \\
bromine               & $\text{Br}_2$              & 10.51 & 6.76    & 9.33     & 9.61      & 6.76     & 9.64      & 9.81       & 6.76      & 10.04      & 10.03       \\
methane               & C$\text{H}_4$              & 14.35 & 14.28   & 13.38    & 13.65     & 14.28    & 13.61     & 13.79      & 14.28     & 14.06      & 14.06       \\
ethane                & $\text{C}_2\text{H}_6$     & 12.20 & 12.56   & 11.55    & 11.85     & 12.56    & 11.81     & 12.01      & 12.56     & 12.29      & 12.30       \\
ethylene              & $\text{C}_2\text{H}_4$     & 10.68 & 10.20   & 9.50     & 9.71      & 10.20    & 9.65      & 9.80       & 10.20     & 9.95       & 9.96        \\
ethyn                 & $\text{C}_2\text{H}_2$     & 11.49 & 11.13   & 10.46    & 10.66     & 11.13    & 10.59     & 10.74      & 11.13     & 10.89      & 10.91       \\
tetracarbon           & $\text{C}_4$               & 12.54 & 10.97   & 9.68     & 10.17     & 10.97    & 10.16     & 10.38      & 10.97     & 10.70      & 10.63       \\
cyclopropane          & $\text{C}_3\text{H}_6$     & 10.54 & 10.54   & 9.34     & 9.70      & 10.54    & 9.65      & 9.89       & 10.54     & 10.19      & 10.20       \\
vinyl fluoride        & $\text{C}_2\text{H}_3$F    & 10.63 & 10.21   & 9.22     & 9.50      & 10.21    & 9.42      & 9.62       & 10.21     & 9.87       & 9.87        \\
vinyl bromide         & $\text{C}_2\text{H}_3$Br   & 9.90  & 8.88    & 8.04     & 8.28      & 8.88     & 8.24      & 8.41       & 8.88      & 8.55       & 8.58        \\
tetrafluoromethane    & C$\text{F}_4$              & 16.20 & 16.16   & 14.00    & 14.74     & 16.16    & 14.62     & 15.13      & 16.16     & 15.91      & 15.93       \\
silane                & Si$\text{H}_4$             & 12.82 & 12.90   & 12.12    & 12.36     & 12.90    & 12.31     & 12.47      & 12.90     & 12.63      & 12.66       \\
germane               & Ge$\text{H}_4$             & 12.46 & 12.54   & 11.76    & 11.99     & 12.54    & 11.94     & 12.10      & 12.54     & 12.25      & 12.28       \\
lithium hydride       & LiH                        & 7.90  & 8.14    & 7.98     & 8.04      & 8.14     & 8.04      & 8.08       & 8.14      & 8.22       & 8.18        \\
potassium hydride     & KH                         & 8.00  & 6.33    & 6.13     & 6.22      & 6.33     & 6.23      & 6.28       & 6.33      & 6.47       & 6.41        \\
borane                & B$\text{H}_3$              & 12.03 & 13.30   & 12.69    & 12.87     & 13.30    & 12.83     & 12.96      & 13.30     & 13.13      & 13.14       \\
diborane              & $\text{B}_2\text{H}_6$     & 11.90 & 12.32   & 11.31    & 11.62     & 12.32    & 11.55     & 11.76      & 12.32     & 11.98      & 12.01       \\
ammonia               & N$\text{H}_3$              & 10.82 & 10.66   & 9.59     & 9.92      & 10.66    & 9.88      & 10.10      & 10.66     & 10.50      & 10.47       \\
hydrazoic acid        & H$\text{N}_3$              & 10.72 & 10.46   & 9.31     & 9.68      & 10.46    & 9.61      & 9.86       & 10.46     & 10.08      & 10.13       \\
phosphine             & P$\text{H}_3$              & 10.59 & 10.38   & 9.73     & 9.90      & 10.38    & 9.87      & 9.99       & 10.38     & 10.11      & 10.13       \\
arsine                & As$\text{H}_3$             & 10.58 & 10.15   & 9.57     & 9.71      & 10.15    & 9.70      & 9.79       & 10.15     & 9.92       & 9.91        \\
hydrogen sulfide      & S$\text{H}_2$              & 10.50 & 10.19   & 9.56     & 9.72      & 10.19    & 9.72      & 9.82       & 10.19     & 9.98       & 9.97        \\
hydrogen fluoride     & FH                         & 16.12 & 15.69   & 14.48    & 14.88     & 15.69    & 14.90     & 15.15      & 15.69     & 15.92      & 15.79       \\
hydrogen chloride     & ClH                        & 12.79 & 12.51   & 11.83    & 12.01     & 12.51    & 12.02     & 12.13      & 12.51     & 12.35      & 12.32       \\
lithium fluoride      & LiF                        & 11.30 & 10.91   & 9.91     & 10.27     & 10.91    & 10.27     & 10.50      & 10.91     & 11.47      & 11.24       \\
magnesium fluoride    & $\text{F}_2$Mg             & 13.30 & 13.35   & 12.05    & 12.55     & 13.35    & 12.48     & 12.80      & 13.35     & 13.78      & 13.59       \\
aluminum fluoride     & Al$\text{F}_3$             & 15.45 & 15.08   & 13.45    & 14.04     & 15.08    & 13.92     & 14.33      & 15.08     & 15.22      & 15.12       \\
boron monofluoride    & BF                         & 11.00 & 10.93   & 10.25    & 10.47     & 10.93    & 10.35     & 10.52      & 10.93     & 10.58      & 10.65       \\
gallium monochloride  & GaCl                       & 10.07 & 9.43    & 8.84     & 9.00      & 9.43     & 8.96      & 9.07       & 9.43      & 9.22       & 9.20        \\
sodium chloride       & NaCl                       & 9.80  & 8.99    & 8.35     & 8.55      & 8.99     & 8.57      & 8.68       & 8.99      & 9.07       & 8.95        \\
magnesium chloride    & Mg$\text{Cl}_2$            & 11.80 & 11.64   & 10.76    & 11.05     & 11.64    & 11.04     & 11.21      & 11.64     & 11.52      & 11.49       \\
boron nitride         & BN                         &       & 11.13   & 10.24    & 10.59     & 11.13    & 10.63     & 10.83      & 11.13     & 11.18      & 11.14       \\
hydrogen cyanide      & NCH                        & 13.61 & 13.48   & 12.80    & 13.01     & 13.48    & 12.96     & 13.11      & 13.48     & 13.30      & 13.30       \\
phosphorus mononitr   & PN                         & 11.88 & 11.93   & 9.48     & 10.20     & 11.93    & 10.15     & 10.61      & 11.93     & 10.96      & 11.06       \\
hydrazine             & $\text{H}_2$NN$\text{H}_2$ & 8.98  & 9.46    & 8.17     & 8.57      & 9.46     & 8.55      & 8.80       & 9.46      & 9.22       & 9.21        \\
formaldehyde          & $\text{H}_2$CO             & 10.89 & 10.62   & 9.18     & 9.64      & 10.62    & 9.65      & 9.93       & 10.62     & 10.41      & 10.39       \\
methanol              & C$\text{H}_2$O             & 10.96 & 10.86   & 9.39     & 9.85      & 10.86    & 9.82      & 10.11      & 10.86     & 10.66      & 10.61       \\
ethanol               & $\text{C}_2\text{H}_6$O    & 10.64 & 10.53   & 8.90     & 9.41      & 10.53    & 9.38      & 9.69       & 10.53     & 10.26      & 10.21       \\
acetaldehyde          & $\text{C}_2\text{H}_4$O    & 10.24 & 10.03   & 8.35     & 8.89      & 10.03    & 8.89      & 9.21       & 10.03     & 9.71       & 9.70        \\
formic acid           & C$\text{H}_2\text{O}_2$    & 11.50 & 11.22   & 9.54     & 10.11     & 11.22    & 10.07     & 10.43      & 11.22     & 10.99      & 10.99       \\
hydrogen peroxide     & HOOH                       & 11.70 & 11.23   & 9.45     & 10.04     & 11.23    & 10.01     & 10.37      & 11.23     & 11.05      & 11.02       \\
water                 & $\text{H}_2$O              & 12.62 & 12.30   & 11.11    & 11.49     & 12.30    & 11.48     & 11.72      & 12.30     & 12.32      & 12.24       \\
carbon dioxide        & C$\text{O}_2$              & 13.77 & 13.57   & 12.00    & 12.52     & 13.57    & 12.41     & 12.80      & 13.57     & 13.24      & 13.27       \\
carbon oxide sulfide  & OCS                        & 11.19 & 11.09   & 10.03    & 10.36     & 11.09    & 10.29     & 10.52      & 11.09     & 10.66      & 10.72       \\
carbon oxide selenide & OCSe                       & 10.37 & 10.28   & 9.42     & 9.67      & 10.28    & 9.62      & 9.80       & 10.28     & 9.92       & 9.96        \\
carbon monoxide       & CO                         & 14.01 & 14.47   & 12.63    & 13.20     & 14.47    & 13.04     & 13.45      & 14.47     & 13.67      & 13.83       \\
ozone                 & $\text{O}_3$               & 12.73 & 12.94   & 9.52     & 10.48     & 12.94    & 10.43     & 11.04      & 12.94     & 11.71      & 11.85       \\
sulfur dioxide        & S$\text{O}_2$              & 12.50 & 11.38   & 11.88    & 11.75     & 10.07    & 11.85     & 11.71      & 8.45      & 11.76      & 11.39       \\
beryllium monoxide    & BeO                        & 10.10 & 9.22    & 8.37     & 8.72      & 9.22     & 8.77      & 8.95       & 9.22      & 9.77       & 9.53        \\
magnesium monoxide    & MgO                        & 8.76  & 7.85    & 5.10     & 5.68      & 7.85     & 5.75      & 6.09       & 7.85      & 7.26       & 7.00        \\
copper dimer          & $\text{Cu}_2$              & 7.46  & 6.88    & 4.87     & 6.93      & 6.88     & 5.67      & 6.96       &           & 7.97       & 7.16        \\
copper cyanide        & NCCu                       &       & 10.82   & 7.79     & 8.54      & 10.82    & 8.42      & 8.89       & 10.82     & 9.90       & 9.75   
\end{longtable}
}

\fontsize{9}{11}\selectfont{
\begin{longtable}{@{\extracolsep{\fill}}llcccccccccc}
\caption{Electron affinities of the GW100 set obtained with def2-TZVPP basis set from $G_{0}T_{0}$, $G_{\text{RS}}T_0$ and $G_{\text{RS}}T_{\text{RS}}$ with different starting points including HF, PBE and PBE0. All values in eV. }\\\toprule
 & & & \multicolumn{3}{c}{$G_{0}T_{0}$} & \multicolumn{3}{c}{$G_{\text{RS}}T_0$}  & \multicolumn{3}{c}{$G_{\text{RS}}T_{\text{RS}}$} \\
 \cmidrule(l{0.5em}r{0.5em}){4-6} \cmidrule(l{0.5em}r{0.5em}){7-9} \cmidrule(l{0.5em}r{0.5em}){10-12}
name & formula & Exp & HF & PBE & PBE0 & HF & PBE & PBE0 & HF & PBE & PBE0  \\
\hline 
\endfirsthead
\caption{Continued}\\\toprule
 & & & \multicolumn{3}{c}{$G_{0}T_{0}$} & \multicolumn{3}{c}{$G_{\text{RS}}T_0$}  & \multicolumn{3}{c}{$G_{\text{RS}}T_{\text{RS}}$} \\
 \cmidrule(l{0.5em}r{0.5em}){4-6} \cmidrule(l{0.5em}r{0.5em}){7-9} \cmidrule(l{0.5em}r{0.5em}){10-12}
name & formula & Exp & HF & PBE & PBE0 & HF & PBE & PBE0 & HF & PBE & PBE0  \\
\hline
\endhead
\bottomrule
\endlastfoot
helium                & He                         &      & -21.75 & -21.63 & -21.66 & -21.75 & -21.67 & -21.69 & -21.75 & -21.72 & -21.72 \\
neon                  & Ne                         &      & -19.58 & -18.99 & -19.19 & -19.58 & -19.16 & -19.29 & -19.58 & -19.31 & -19.40 \\
argon                 & Ar                         &      & -10.49 & -10.00 & -10.18 & -10.49 & -10.20 & -10.30 & -10.49 & -10.47 & -10.50 \\
krypton               & Kr                         &      & -9.42  & -9.24  & -9.32  & -9.42  & -9.33  & -9.37  & -9.42  & -9.46  & -9.46  \\
hydrogen              & $\text{H}_2$               &      & -4.23  & -4.26  & -4.24  & -4.23  & -4.33  & -4.27  & -4.23  & -4.41  & -4.33  \\
lithium dimer         & $\text{Li}_2$              &      & -0.07  & 0.07   & 0.00   & -0.07  & -0.05  & -0.06  & -0.07  & -0.25  & -0.19  \\
sodium dimer          & $\text{Na}_2$              & 0.54 & 0.15   & 0.20   & 0.17   & 0.15   & 0.11   & 0.12   & 0.15   & -0.06  & 0.01   \\
sodium tetramer       & $\text{Na}_4$              & 0.91 & 0.43   & 0.72   & 0.57   & 0.43   & 0.57   & 0.50   & 0.43   & 0.28   & 0.31   \\
potassium dimer       & $\text{K}_2$               & 0.50 & 0.25   & 0.34   & 0.29   & 0.25   & 0.24   & 0.24   & 0.25   & 0.06   & 0.12   \\
nitrogen              & $\text{N}_2$               &      & -2.84  & -2.27  & -2.49  & -2.84  & -2.56  & -2.64  & -2.84  & -3.11  & -3.02  \\
phosphorus dimer      & $\text{P}_2$               & 0.63 & 0.42   & 0.80   & 0.63   & 0.42   & 0.63   & 0.55   & 0.42   & 0.15   & 0.19   \\
arsenic dimer         & $\text{As}_2$              & 0.74 & 4.35   & 0.81   & 0.67   & 4.35   & 0.68   & 0.61   & 4.35   & 0.26   & 0.30   \\
fluorine              & $\text{F}_2$               &      & -0.05  & 1.70   & 1.13   & -0.05  & 1.16   & 0.82   & -0.05  & 0.48   & 0.34   \\
chlorine              & $\text{Cl}_2$              &      & 0.43   & 1.19   & 0.90   & 0.43   & 0.90   & 0.75   & 0.43   & 0.36   & 0.35   \\
bromine               & $\text{Br}_2$              &      & 7.10   & 1.63   & 1.39   & 7.10   & 1.40   & 1.26   & 7.10   & 0.97   & 0.94   \\
methane               & C$\text{H}_4$              &      & -3.49  & -3.22  & -3.31  & -3.49  & -3.37  & -3.39  & -3.49  & -3.52  & -3.50  \\
ethane                & $\text{C}_2\text{H}_6$     &      & -3.15  & -2.78  & -2.91  & -3.15  & -2.96  & -3.01  & -3.15  & -3.16  & -3.14  \\
ethylene              & $\text{C}_2\text{H}_4$     &      & -2.60  & -1.80  & -2.09  & -2.60  & -2.14  & -2.27  & -2.60  & -2.63  & -2.60  \\
ethyn                 & $\text{C}_2\text{H}_2$     &      & -3.64  & -2.89  & -3.09  & -3.64  & -3.16  & -3.22  & -3.64  & -3.60  & -3.52  \\
tetracarbon           & $\text{C}_4$               & 3.88 & 2.47   & 3.44   & 3.09   & 2.47   & 3.04   & 2.87   & 2.47   & 2.30   & 2.36   \\
cyclopropane          & $\text{C}_3\text{H}_6$     &      & -3.49  & -2.99  & -3.17  & -3.49  & -3.22  & -3.28  & -3.49  & -3.48  & -3.46  \\
vinyl fluoride        & $\text{C}_2\text{H}_3$F    &      & -2.72  & -1.99  & -2.26  & -2.72  & -2.36  & -2.47  & -2.72  & -2.82  & -2.79  \\
vinyl bromide         & $\text{C}_2\text{H}_3$Br   &      & -1.89  & -1.03  & -1.35  & -1.89  & -1.39  & -1.54  & -1.89  & -1.93  & -1.93  \\
tetrafluoromethane    & C$\text{F}_4$              &      & -4.92  & -4.31  & -4.51  & -4.92  & -4.65  & -4.67  & -4.92  & -4.93  & -4.84  \\
silane                & Si$\text{H}_4$             &      & -3.21  & -2.91  & -3.02  & -3.21  & -3.11  & -3.13  & -3.21  & -3.36  & -3.29  \\
germane               & Ge$\text{H}_4$             &      & -3.09  & -2.71  & -2.83  & -3.09  & -2.91  & -2.94  & -3.09  & -3.14  & -3.10  \\
lithium hydride       & LiH                        & 0.34 & -0.13  & -0.28  & -0.22  & -0.13  & -0.34  & -0.25  & -0.13  & -0.38  & -0.28  \\
potassium hydride     & KH                         &      & 0.01   & -0.12  & -0.09  & 0.01   & -0.20  & -0.12  & 0.01   & -0.26  & -0.16  \\
borane                & B$\text{H}_3$              & 0.04 & -0.63  & -0.17  & -0.33  & -0.63  & -0.49  & -0.50  & -0.63  & -0.84  & -0.73  \\
diborane              & $\text{B}_2\text{H}_6$     &      & -1.44  & -0.70  & -0.95  & -1.44  & -1.05  & -1.14  & -1.44  & -1.46  & -1.42  \\
ammonia               & N$\text{H}_3$              &      & -2.89  & -2.65  & -2.72  & -2.89  & -2.80  & -2.81  & -2.89  & -2.94  & -2.90  \\
hydrazoic acid        & H$\text{N}_3$              &      & -1.82  & -1.05  & -1.29  & -1.82  & -1.39  & -1.48  & -1.82  & -2.03  & -1.93  \\
phosphine             & P$\text{H}_3$              &      & -2.97  & -2.74  & -2.90  & -2.97  & -2.96  & -3.01  & -2.97  & -3.28  & -3.23  \\
arsine                & As$\text{H}_3$             &      & -2.88  & -2.73  & -2.88  & -2.88  & -2.92  & -2.98  & -2.88  & -3.21  & -3.19  \\
hydrogen sulfide      & S$\text{H}_2$              &      & -2.74  & -2.75  & -2.91  & -2.74  & -2.97  & -3.03  & -2.74  & -3.31  & -3.26  \\
hydrogen fluoride     & FH                         &      & -3.05  & -3.02  & -3.00  & -3.05  & -3.14  & -3.07  & -3.05  & -3.24  & -3.13  \\
hydrogen chloride     & ClH                        &      & -2.60  & -2.41  & -2.46  & -2.60  & -2.60  & -2.56  & -2.60  & -2.88  & -2.75  \\
lithium fluoride      & LiF                        &      & 0.01   & -0.24  & -0.12  & 0.01   & -0.28  & -0.15  & 0.01   & -0.30  & -0.16  \\
magnesium fluoride    & $\text{F}_2$Mg             &      & 0.07   & 0.05   & 0.02   & 0.07   & -0.10  & -0.05  & 0.07   & -0.16  & -0.10  \\
aluminum fluoride     & Al$\text{F}_3$             &      & -0.43  & -0.17  & -0.29  & -0.43  & -0.42  & -0.42  & -0.43  & -0.56  & -0.52  \\
boron monofluoride    & BF                         &      & -1.61  & -1.49  & -1.54  & -1.61  & -1.75  & -1.67  & -1.61  & -2.05  & -1.87  \\
gallium monochloride  & GaCl                       &      & -0.26  & -0.10  & -0.20  & -0.26  & -0.31  & -0.31  & -0.26  & -0.57  & -0.49  \\
sodium chloride       & NaCl                       & 0.73 & 0.57   & 0.31   & 0.41   & 0.57   & 0.23   & 0.37   & 0.57   & 0.18   & 0.34   \\
magnesium chloride    & Mg$\text{Cl}_2$            &      & 0.20   & 0.37   & 0.29   & 0.20   & 0.17   & 0.19   & 0.20   & -0.04  & 0.04   \\
boron nitride         & BN                         & 3.16 & 3.93   & 3.95   & 3.84   & 3.93   & 3.77   & 3.75   & 3.93   & 3.27   & 3.41   \\
hydrogen cyanide      & NCH                        &      & -3.40  & -2.54  & -2.75  & -3.40  & -2.83  & -2.90  & -3.40  & -3.31  & -3.22  \\
phosphorus mononitr   & PN                         &      & -0.15  & 0.33   & 0.12   & -0.15  & 0.11   & 0.01   & -0.15  & -0.40  & -0.34  \\
hydrazine             & $\text{H}_2$NN$\text{H}_2$ &      & -2.56  & -2.19  & -2.31  & -2.56  & -2.38  & -2.41  & -2.56  & -2.59  & -2.55  \\
formaldehyde          & $\text{H}_2$CO             &      & -1.62  & -0.66  & -1.01  & -1.62  & -1.07  & -1.22  & -1.62  & -1.60  & -1.59  \\
methanol              & C$\text{H}_2$O             &      & -3.09  & -2.68  & -2.82  & -3.09  & -2.90  & -2.93  & -3.09  & -3.11  & -3.07  \\
ethanol               & $\text{C}_2\text{H}_6$O    &      & -2.93  & -2.46  & -2.62  & -2.93  & -2.69  & -2.74  & -2.93  & -2.93  & -2.89  \\
acetaldehyde          & $\text{C}_2\text{H}_4$O    &      & -1.88  & -0.70  & -1.13  & -1.88  & -1.16  & -1.37  & -1.88  & -1.76  & -1.79  \\
formic acid           & C$\text{H}_2\text{O}_2$    &      & -3.13  & -1.71  & -2.06  & -3.13  & -2.13  & -2.29  & -3.13  & -2.64  & -2.65  \\
hydrogen peroxide     & HOOH                       &      & -3.04  & -1.52  & -2.07  & -3.04  & -2.13  & -2.38  & -3.04  & -2.76  & -2.78  \\
water                 & $\text{H}_2$O              &      & -2.90  & -2.70  & -2.75  & -2.90  & -2.85  & -2.83  & -2.90  & -2.98  & -2.92  \\
carbon dioxide        & C$\text{O}_2$              &      & -2.80  & -3.64  & -2.57  & -2.80  & -3.97  & -2.71  & -2.80  & -4.42  & -2.90  \\
carbon oxide sulfide  & OCS                        & 0.46 & -1.54  & -1.00  & -1.25  & -1.54  & -1.31  & -1.41  & -1.54  & -1.91  & -1.84  \\
carbon oxide selenide & OCSe                       &      & -1.15  & -0.58  & -0.84  & -1.15  & -0.92  & -1.01  & -1.15  & -1.55  & -1.47  \\
carbon monoxide       & CO                         & 1.33 & -1.03  & -0.67  & -0.83  & -1.03  & -0.95  & -0.97  & -1.03  & -1.46  & -1.32  \\
ozone                 & $\text{O}_3$               & 2.10 & 2.44   & 2.82   & 2.63   & 2.44   & 2.64   & 2.55   & 2.44   & 2.14   & 2.19   \\
sulfur dioxide        & S$\text{O}_2$              & 1.11 & 6.30   & 6.77   & 6.65   & 12.05  & 3.91   & 5.15   & 6.86   & 7.19   & 7.10   \\
beryllium monoxide    & BeO                        &      & 2.05   & 1.82   & 1.86   & 2.05   & 1.65   & 1.77   & 2.05   & 1.50   & 1.67   \\
magnesium monoxide    & MgO                        &      & 1.60   & 1.91   & 1.74   & 1.60   & 1.73   & 1.66   & 1.60   & 1.58   & 1.56   \\
copper dimer          & $\text{Cu}_2$              & 0.84 & 0.15   & 1.02   & 0.63   & 0.15   & 0.51   & 0.46   & 0.15   & 0.34   & 0.29   \\
copper cyanide        & NCCu                       & 1.47 & 0.82   & 1.85   & 1.36   & 0.82   & 1.40   & 1.15   & 0.82   & 1.12   & 0.98  
\end{longtable}
}

\section{CORE65 set results}
The CLBEs of molecules in the CORE65\cite{golze2019accurate} set calculated from $G_{0}T_{0}$, $G_{\text{RS}}T_0$ and $G_{\text{RS}}T_{\text{RS}}$ based on PBE, B3LYP, PBE0 and HF with def2-TZVP are presented in this section. The reference values are taken from the Golze's work\cite{golze2019accurate}. 8 molecules are excluded considering computational cost. All calculations are performed by our software QM4D.

\begingroup
\setlength{\tabcolsep}{4pt}
\renewcommand{\arraystretch}{1}
 \fontsize{9}{11}\selectfont{
\begin{longtable}{llcccccccccccc}
\caption{Core level binding energies of the CORE65 set obtained with def2-TZVP basis set from $G_{0}T_{0}$, $G_{\text{RS}}T_0$ and $G_{\text{RS}}T_{\text{RS}}$ with different starting points including HF, PBE, PBE0, B3LYP. All values in eV.}\\\toprule
 & & \multicolumn{4}{c}{$G_{0}T_{0}$} & \multicolumn{4}{c}{$G_{\text{RS}}T_0$}  & \multicolumn{4}{c}{$G_{\text{RS}}T_{\text{RS}}$} \\
 \cmidrule(l{0.5em}r{0.5em}){3-6} \cmidrule(l{0.5em}r{0.5em}){7-10} \cmidrule(l{0.5em}r{0.5em}){11-14}
formula & core level & PBE & PBE0 & B3LYP & HF & PBE & PBE0 & B3LYP & HF & PBE & PBE0 & B3LYP & HF  \\
\hline 
\endfirsthead
\caption{Continued}\\\toprule
 & & \multicolumn{4}{c}{$G_{0}T_{0}$} & \multicolumn{4}{c}{$G_{\text{RS}}T_0$}  & \multicolumn{4}{c}{$G_{\text{RS}}T_{\text{RS}}$} \\
 \cmidrule(l{0.5em}r{0.5em}){3-6} \cmidrule(l{0.5em}r{0.5em}){7-10} \cmidrule(l{0.5em}r{0.5em}){11-14}
formula & core level & PBE & PBE0 & B3LYP & HF & PBE & PBE0 & B3LYP & HF & PBE & PBE0 & B3LYP & HF \\
\hline
\endhead
\bottomrule
\endlastfoot
C$\text{H}_4$                 & C1s                  & 280.14 & 285.51 & 284.50 & 294.76 & 282.19 & 286.74 & 285.92 & 294.76 & 292.07 & 293.28 & 292.26 & 294.76 \\
$\text{C}_2\text{H}_6$        & C1s                  & 279.44 & 285.01 & 283.97 & 294.51 & 281.58 & 286.35 & 285.49 & 294.51 & 292.39 & 292.78 & 292.51 & 294.51 \\
$\text{C}_2\text{H}_4$        & C1s                  & 278.85 & 284.86 & 283.76 & 294.93 & 281.33 & 286.32 & 285.43 & 294.93 & 292.69 & 293.12 & 292.82 & 294.93 \\
$\text{C}_2\text{H}_2$        & C1s                  & 279.01 & 285.07 & 283.95 & 295.16 & 281.14 & 286.15 & 285.24 & 295.16 & 293.01 & 293.48 & 293.15 & 295.16 \\
CO                            & O1s                  & 524.51 & 532.27 & 530.70 &        & 527.49 & 533.92 & 532.62 & 545.10 & 543.65 & 544.00 & 543.50 & 545.10 \\
CO                            & C1s                  & 285.49 & 291.50 & 290.50 &        & 288.05 & 292.87 & 292.10 & 300.57 & 298.14 & 298.71 & 298.48 & 300.57 \\
C$\text{O}_2$                 & O1s                  & 522.05 & 530.59 & 528.92 & 544.49 & 525.02 & 532.41 & 531.01 & 544.49 & 542.36 & 542.94 & 542.40 & 544.49 \\
C$\text{O}_2$                 & C1s                  & 286.19 & 292.24 & 291.32 & 302.69 & 288.61 & 293.70 & 293.01 & 302.69 & 298.75 & 299.71 & 299.55 & 302.69 \\
C$\text{F}_4$                 & F1s                  & 675.77 & 684.35 & 682.86 & 698.41 & 678.86 & 686.29 & 684.94 & 698.41 & 696.77 & 697.25 & 696.81 & 698.41 \\
C$\text{F}_4$                 & C1s                  & 289.84 & 295.84 & 295.04 & 306.34 & 292.26 & 297.36 & 296.69 & 306.34 & 302.61 & 303.53 & 303.40 & 306.34 \\
CF$\text{H}_3$                & F1s                  & 673.42 & 681.79 & 680.12 & 695.36 & 676.24 & 683.47 & 682.07 & 695.36 & 694.27 & 694.51 & 694.01 & 695.36 \\
CF$\text{H}_3$                & C1s                  & 282.67 & 288.27 & 287.29 & 297.65 & 284.79 & 289.51 & 288.73 & 297.65 & 295.43 & 295.90 & 295.67 & 297.65 \\
C$\text{F}_3$H                & F1s                  & 674.87 & 683.55 & 682.05 & 697.46 & 677.85 & 685.32 & 683.92 & 697.46 & 696.04 & 696.36 & 695.91 & 697.46 \\
C$\text{F}_3$H                & C1s                  & 287.65 & 293.60 & 292.67 & 303.58 & 290.00 & 294.92 & 294.22 & 303.58 & 300.36 & 301.13 & 300.97 & 303.58 \\
C$\text{H}_3$OH               & O1s                  & 521.88 & 529.63 & 527.91 & 542.04 & 524.11 & 531.08 & 529.78 & 542.04 & 540.43 & 540.80 & 540.28 & 542.04 \\
C$\text{H}_3$OH               & C1s                  & 281.36 & 286.95 & 285.95 & 296.42 & 283.55 & 288.24 & 287.43 & 296.42 & 294.26 & 294.70 & 294.45 & 296.42 \\
C$\text{H}_2$O                & O1s                  & 521.82 & 529.09 & 527.47 & 542.40 & 524.14 & 530.85 & 529.47 & 542.40 & 540.87 & 541.25 & 540.70 & 542.40 \\
C$\text{H}_2$O                & C1s                  & 283.37 & 289.23 & 288.25 & 298.86 & 285.82 & 290.67 & 289.88 & 298.86 & 296.40 & 296.93 & 296.71 & 298.86 \\
C$\text{H}_3$-O-C$\text{H}_3$ & O1s                  & 520.71 & 528.26 & 526.52 & 541.56 & 523.18 & 529.98 & 528.57 & 541.56 & 540.04 & 540.41 & 539.82 & 541.56 \\
C$\text{H}_3$-O-C$\text{H}_3$ & C1s                  & 280.84 & 286.56 & 285.58 & 296.25 & 283.08 & 287.67 & 287.07 & 296.25 & 294.00 & 294.47 & 294.22 & 296.25 \\
HCOOH                         & O1s\_OH              & 522.46 & 530.40 & 528.64 & 543.92 & 525.21 & 531.93 & 530.71 & 543.92 & 542.07 & 542.52 & 541.99 & 543.92 \\
HCOOH                         & O1s\_C=O             & 520.13 & 528.49 & 526.86 & 542.07 & 523.16 & 530.20 & 528.83 & 542.07 & 540.30 & 540.75 & 540.23 & 542.07 \\
HCOOH                         & C1s                  & 284.17 & 290.19 & 289.22 & 300.24 & 286.54 & 291.54 & 290.79 & 300.24 & 297.19 & 297.92 & 297.71 & 300.24 \\
C$\text{H}_3$C$\text{O}_2$H   & O1s\_OC$\text{H}_3$  & 520.74 & 529.23 & 527.53 & 543.04 & 523.92 & 530.99 & 529.55 & 543.04 & 541.23 & 541.66 & 541.09 & 543.04 \\
C$\text{H}_3$C$\text{O}_2$H   & O1s\_C=O             & 519.57 & 528.07 & 526.42 & 541.75 & 522.71 & 529.83 & 528.44 & 541.75 & 539.94 & 540.41 & 539.87 & 541.75 \\
C$\text{H}_3$COOH             & O1s\_OH              & 521.94 & 529.94 & 528.20 & 543.51 & 524.58 & 531.63 & 530.23 & 543.51 & 541.54 & 542.01 & 541.47 & 543.51 \\
C$\text{H}_3$COOH             & O1s\_C=O             & 519.51 & 527.86 & 526.23 & 541.50 & 522.40 & 529.52 & 528.14 & 541.50 & 539.68 & 540.15 & 539.61 & 541.50 \\
C$\text{H}_3$COOH             & C1s\_COOH            & 283.10 & 289.28 & 288.24 & 299.71 & 285.47 & 290.65 & 289.84 & 299.71 & 296.59 & 297.33 & 297.09 & 299.71 \\
C$\text{H}_3$COOH             & C1s\_C$\text{H}_3$   & 280.06 & 285.92 & 284.89 & 295.42 & 282.47 & 287.19 & 286.49 & 295.42 & 293.41 & 293.83 & 293.56 & 295.42 \\
$\text{H}_2$O                 & O1s                  & 523.58 & 530.69 & 529.16 & 542.88 & 526.00 & 532.11 & 530.84 & 542.88 & 541.42 & 541.71 & 541.27 & 542.88 \\
$\text{O}_3$                  & O1s\_midd            & 527.59 & 536.00 &        & 551.08 & 530.76 & 537.90 & 536.46 & 551.08 & 547.90 & 548.66 & 548.06 & 551.08 \\
$\text{O}_3$                  & O1s\_term            & 519.94 & 529.47 &        & 545.42 & 523.32 & 531.41 & 529.99 & 545.42 & 542.50 & 543.23 & 542.77 & 545.42 \\
$\text{N}_2$                  & N1s                  & 394.81 & 401.92 & 400.53 & 413.51 & 397.43 & 402.16 & 401.31 & 413.51 & 411.21 & 411.76 & 411.37 & 413.51 \\
N$\text{H}_3$                 & N1s                  & 391.85 & 398.21 & 396.86 & 409.17 & 394.09 & 399.54 & 398.41 & 409.17 & 407.30 & 407.70 & 407.25 & 409.17 \\
HCN                           & N1s                  & 390.82 & 398.13 & 396.64 & 410.43 & 393.63 & 399.73 & 398.52 & 410.43 & 408.28 & 408.78 & 408.33 & 410.43 \\
HCN                           & C1s                  & 281.89 & 287.72 & 286.70 & 297.22 & 283.73 & 289.26 & 288.43 & 297.22 & 295.04 & 295.52 & 295.26 & 297.22 \\
C$\text{H}_3$CN               & N1s                  & 389.30 & 396.69 & 395.01 & 409.42 & 392.06 & 398.53 & 397.31 & 409.42 & 407.24 & 407.80 & 407.38 & 409.42 \\
C$\text{H}_3$CN               & C1s\_C$\text{H}_3$   & 280.63 & 286.83 & 285.77 & 296.64 & 282.96 & 287.98 & 287.39 & 296.64 & 294.38 & 294.84 & 294.55 & 296.64 \\
C$\text{H}_3$CN               & C1s\_CN              & 280.69 & 286.62 & 285.56 & 296.41 & 282.96 & 287.97 & 287.14 & 296.41 & 294.17 & 294.71 & 294.42 & 296.41 \\
CO$(\text{NH}_2)_2$           & O1s                  & 518.40 & 526.76 & 525.12 & 540.45 & 521.19 & 528.39 & 526.99 & 540.45 & 538.59 & 539.05 & 538.51 & 540.45 \\
CO$(\text{NH}_2)_2$           & N1s                  & 391.14 & 398.23 & 396.80 & 409.88 & 393.69 & 399.69 & 398.49 & 409.88 & 407.78 & 408.28 & 407.79 & 409.88 \\
CO$(\text{NH}_2)_2$           & C1s                  & 282.63 & 288.74 & 287.70 & 299.29 & 284.95 & 290.12 & 289.32 & 299.29 & 296.02 & 296.78 & 296.55 & 299.29 \\
C$\text{H}_3\text{NH}_2$      & N1s                  & 390.77 & 397.41 & 396.00 & 408.71 & 393.10 & 398.81 & 397.61 & 408.71 & 406.85 & 407.28 & 406.78 & 408.71
\end{longtable}
}
\endgroup

\bibliography{ref.bib}